\documentclass[aps,prc,preprint,groupedaddress]{revtex4-2}
\usepackage[utf8]{inputenc}
\usepackage{amsmath}
\usepackage[]{amssymb}
\usepackage[]{mathrsfs}
\usepackage{xcolor}
\usepackage{graphicx}
\usepackage{float}
\usepackage{array}
\usepackage[T1]{fontenc}
\usepackage[normalem]{ulem}

\begin{document}

\title{ Subtracted Second Tamm-Dancoff Approximation in the relativistic point-coupling model}

\date{\today}

\author{Deni Vale}
\affiliation{Istrian University of Applied Sciences, Preradovićeva 9D, HR-52100 Pula, Croatia}

\author{Nils Paar}
\affiliation{Department of Physics, Faculty of Science, University of Zagreb, Bijenička cesta 32, HR-10000 Zagreb, Croatia}

\begin{abstract}
Theoretical description of collective nuclear excitations and astrophysically relevant processes require methods going beyond the Random Phase Approximation (RPA) or Tamm-Dancoff Approximation (TDA), which are limited to one-particle–one-hole (1p-1h) configurations. In this work, we introduce the Second Tamm-Dancoff Approximation (STDA) based on the relativistic nuclear energy density functional,
which also incorporates two-particle–two-hole (2p-2h) configurations, providing a more comprehensive description of nuclear excitations.  
Within the relativistic STDA (RSTDA), we implement the subtraction method, that is essential to avoid double counting of correlations and provide realistic description of excited states. 
Using the DD-PC1 parametrization of the relativistic point-coupling interaction, in the first application of the RSTDA we investigate the properties of isoscalar monopole and quadrupole transitions in $^{16}$O. The analysis reveals important role of complex configurations on centroid energies, transition strength distributions, and the fine structure of nuclear response.
By resolving the fragmentation and spreading of excitation strength, the RSTDA enables accurate description of experimental spectra.

\end{abstract}

\maketitle


\section{Introduction}

Nuclear excitation phenomena are essential for understanding a wide range of phenomena both in nuclear physics and astrophysics.
From the synthesis of elements in stars to the understanding of exotic nuclei and their role in stellar nucleosynthesis, the study of nuclear excitations impacts numerous aspects of modern science. One of the primary challenges in this domain is providing an accurate theoretical description of collective multipole excitations, such as giant resonances, which are essential for interpreting experimental observations \cite{Baldwin1947, Harakeh1976, Youngblood1977, Ulrikson1977, Youngblood1997, Lui2001, Youngblood2001, Hartmann2002, Tamii2011, Savran2013, HOWARD2020} and refining theoretical models \cite{Goldhaber1948, Paar2009, Dao,  Niu2015, Lanza2023}, as well as for the description of various astrophysically relevant nuclear processes and reactions \cite{Balasi2015, Langanke2019}. These excitations reveal critical information about nuclear dynamics, also enabling a deeper understanding of nuclear forces. 

The Quasiparticle Random-Phase Approximation (QRPA), which is based on the two-quasiparticle or one-particle–one-hole (1p-1h) model space, has long been a standard tool in research of collective nuclear excitations \cite{Thouless1961, Lemmer1968, Reinhard2014,Paar2003,Paar2004}. It has proven effective in describing a wide range of phenomena, particularly in medium-heavy and heavy nuclei. However, its predictive power is limited as it does not fully account for all the complex interactions necessary for a complete description of collective nuclear excitations. To improve
the accuracy in modeling the properties of excitation phenomena, more sophisticated approaches have been developed. These include models that incorporate additional physical effects, such as the theory of finite Fermi systems, which integrates both the single-particle continuum and the coupling to low-lying collective states \cite{Caurier2005, Donno2011}. Furthermore, models like particle-vibration coupling (PVC) and particle-phonon coupling (PPC) enhance the description of low-lying nuclear states by accounting for interactions between single-particle states and collective phonon excitations \cite{Colo2010,MishevVoronov2010,Niu2012,Niu2015,PhysRevLett.131.082501,PhysRevC.110.064317,PhysRevC.101.044316}. In addition to these, other approaches such as the shell model and various density functional theories, including relativistic and non-relativistic formulations, have contributed significantly to refining the understanding of nuclear dynamics and collective phenomena \cite{Litvinova2007_PLB, Dao, Litvinova2011, Knapp, Vesely:2025}. 
In the relativistic framework, the covariant time-blocking approximation (TBA) and its quasiparticle extension (RQTBA) have been introduced, which incorporate particle-vibration coupling and configurations involving two quasiparticles coupled to collective vibrations, resulting in fragmentation and damping of giant resonances
\cite{Litvinova2008, Litvinova2008b, Litvinova2023}. 
This approach is specific in that it projects out nontrivial time orderings beyond the coupling to a single phonon, thus preventing double counting of correlations while preserving causality and computational tractability. 

Both Second Tamm-Dancoff Approximation (STDA) 
and Second Random Phase Approximation (SRPA) were
previously established in non-relativistic framework. The initial formulations given in Refs. \cite{Sawicki1962,Providencia,Yannouleas,Drozdz1990} have been extended to SRPA based on realistic finite-range interactions \cite{Papakonstantinou2009, Papakonstantinou2010}, Skyrme functionals \cite{AitTahar1993, Gambacurta2011, Minato2016,Vasseur2018,Gambacurta2020,Yang2021,Yang2024}, and the Gogny interaction \cite{Gambacurta2012,Gambacurta2016}. More recently, comparative study of various formalisms and performances of SRPA and other beyond-mean-field approaches
is given in Ref. \cite{Knapp}. There are also extensions going beyond simple SRPA \cite{Takayanagi1988,Gambacurta2006}. 
 Initial SRPA implementations, developed several decades ago, due to computational limitations were restricted to a diagonal approximation, only accounting for 1p-1h and 2p-2h couplings without considering interactions among 2p-2h states. Recent advancements in the SRPA have enabled extensive calculations in closed-shell nuclei, such as $^{16}$O and $^{40}$Ca, achieving a more accurate description of strength distributions, fine structure, and transition densities. While previous studies 
based on the SRPA have largely focused on non-relativistic implementations, to date there is no STDA or SRPA established in the relativistic framework.
Only a preliminary simplified relativistic Second Tamm-Dancoff Approximation (RSTDA) model based on the diagonal approximation was previously developed as an initial step toward incorporating interaction between 2p-2h pairs within this framework \cite{ValePaar2024}.

In this work, we present the first fully self-consistent implementation of the STDA within the framework of relativistic energy density functional theory. We focus on analyzing isoscalar (ISGMR) and isovector giant monopole (IVGMR) and isoscalar (ISGQR) and isovector quadrupole (IVGQR) excitations in $^{16}$O. Despite the increased computational demands, the RSTDA offers improved accuracy and valuable insights into the fragmentation of complex excitation spectra. However, the inclusion of 2p-2h configurations can lead to complications such as double counting and infrared instabilities, as analyzed in previous SRPA studies \cite{Tselyaev2007, Tselyaev2013, Gambacurta2015}. 
To address these issues at the RSTDA level, we introduce the subtraction method in the relativistic framework, giving rise to the subtracted RSTDA (RSSTDA), which ensures the stability of the solutions and avoids double counting of correlations, in analogy with previous non-relativistic SRPA studies \cite{Tselyaev2013, Gambacurta2015}.
The RSSTDA introduced in this work represents the first step toward future development of the full SRPA in the relativistic framework.
In Section \ref{two} we describe the relativistic energy density functional (REDF) used in this work. The RSTDA equations and more details about the formalism, including rearrangement terms, are given in Section \ref{three}. In Section \ref{four} we show the computational results for the strength distribution in {$^{16}$O} for the ISGMR and ISGQR, as well as corresponding moments, centroid energies, widths and the exhaustion of the sum rule. We also give microscopic explanation of the subtraction method for the few representative counterpart eigenstates.

\section{Relativistic energy density functional}\label{two}
In the REDF framework, nucleons are represented as a system of Dirac particles interacting through
(i) the finite-range meson exchange interaction 
and (ii) the contact type of interaction, where the meson propagator is replaced by a zero-range interaction \cite{Nik}. In both approaches, the relativistic Lagrangian is constructed using the fundamental symmetries of QCD (Lorentz covariance, gauge invariance, and chiral symmetry). In this work we employ the formalism of the relativistic point-coupling model, introduced in Ref.~\cite{Niksic2008}, and more details about its numerical implementation are given in Ref.~\cite{Nik}.
The basic building blocks of the REDF are densities and currents that are bilinear in the Dirac spinor fields for nucleons \cite{Bor,Niksic2008,Nik}:
\begin{equation}
\bar{\psi} O_\tau \Gamma \psi, ~~O_\tau \in\left\{1, \tau_i\right\}, \quad \Gamma \in\left\{1, \gamma_\mu, \gamma_5, \gamma_5 \gamma_\mu, \sigma_{\mu v}\right\}.
\end{equation}
where $\tau_i$ are the Pauli matrices in isospin space, and $\Gamma$ represents the Dirac matrices or their products. The determination of the ground state density and energy arises from the self-consistent solution of the relativistic linear single-nucleon Kohn-Sham equations.
To derive those equations it is useful to establish an effective Lagrangian density, written by using an expansion of currents $\bar{\psi} \, O \, \tau \, \Gamma \, \psi$ and their derivatives \cite{Niksic2008}. 
The four-fermion (contact) interaction is given in terms of different isospin-spatial space p-h channels: i) isoscalar-scalar $(\bar{\psi} \psi)(\bar{\psi} \psi)$, ii) isoscalar-vector $(\bar{\psi} \gamma_\mu \psi)(\bar{\psi} \gamma^\mu \psi)$, iii) isovector-scalar  $(\bar{\psi} \vec{\tau}\psi)(\bar{\psi} \vec{\tau}\psi)$ and iv) isovector-vector  $(\bar{\psi} \vec{\tau}\gamma_\mu \psi)(\bar{\psi}\vec{\tau} \gamma^\mu \psi)$. In a similar manner, the derivative terms can also be constructed.
Upon including the free-nucleon and electromagnetic components, the resulting Lagrangian density for the nuclear system is given by the expression \cite{Nik}:
\begin{equation}
\begin{aligned}
\mathcal{L}~=&~\bar{\psi}\left(i\gamma_\mu\partial^\mu - m \right)\psi-\frac{1}{2}\alpha_S(\rho)\left(\bar{\psi}\psi\right)\left(\bar{\psi}\psi\right)\\
&-\frac{1}{2}\alpha_V(\rho)\left(\bar{\psi}\gamma_\mu\psi\right)\left(\bar{\psi}\gamma^\mu\psi\right)-\frac{1}{2}\alpha_{TV}(\rho)\left(\bar{\psi}\vec{\tau}\gamma_\mu\psi\right)\left(\bar{\psi}\vec{\tau}\gamma^\mu\psi\right)
\\
&-\frac{1}{2}\delta_S\partial_\nu\left(\bar{\psi}\psi\right)\partial^\nu\left(\bar{\psi}\psi\right)
-e\bar{\psi}\gamma_\mu A^\mu \frac{1-\tau_3}{2}\psi.
\label{lag}
\end{aligned}
\end{equation}
In this formulation, the density dependence is explicitly included in the case of the isoscalar-scalar $\alpha_S(\rho)$, isoscalar-vector $\alpha_V(\rho)$ and isovector-vector coupling $\alpha_{TV}(\rho)$ as 
\begin{equation}
\alpha_i\left[ x \right]~=~a_i + \left( b_i + c_i x\right)\exp(-d_i x),
\end{equation}
where $x = \rho/\rho_{sat}$, $\rho$ stands for barion (vector) density and $\rho_{sat}$ for the nucleon density at saturation in the symmetric nuclear matter \cite{Nik}. In the calculations in this work we employ the DD-PC1 parameterization for $a_i, ~b_i, ~c_i$ and $d_i$ as given in Ref. \cite{Nik}.

\section{Relativistic subtracted second TDA}\label{three}
In the relativistic form of TDA the excitation operators are assumed to be linear superposition of particle-hole (1p-1h) and antiparticle-hole (1$\alpha$-1h) creation and annihilation operators:
\begin{equation}
Q^{\dagger}_\nu ~=~ \sum_{ph} X_{ph}^\nu a_p^\dagger a_h + \sum_{\alpha h} X_{\alpha h}^\nu a_\alpha^\dagger a_h.
\label{TDAQ}
\end{equation}
For simplicity, the coupling to total angular momentum is not included here. In the case of the relativistic STDA, we need to include 2p-2h, 1 particle–1 antiparticle–2 hole (1p-$\alpha$-2h) and 2 antiparticle–2 hole (2$\alpha$-2h) pairs, in addition to the standard ones from eq. \eqref{TDAQ} \cite{ValePaar2024}:
\begin{equation}
\begin{aligned}
Q^\dagger_\nu ~=&~\sum_{ph} X_{ph}^\nu a_p^\dagger a_h + \sum_{\alpha h} X_{\alpha h}^\nu a_\alpha^\dagger a_h
\\&~+ \sum_{p<p'h<h'} \mathcal{X}_{pp'hh'}^\nu a_p^\dagger a_{p'}^\dagger a_{h'} a_h + \sum_{p\alpha h < h'} \mathcal{X}_{p\alpha h h'}^\nu a_p^\dagger a_{\alpha}^\dagger a_{h'} a_h\\
&~+\sum_{\alpha < \alpha' h  < h'} \mathcal{X}_{\alpha \alpha' h h'}^\nu a_\alpha^\dagger a_{\alpha'}^\dagger a_{h'} a_h.\\
\end{aligned}
\end{equation} 
The energies $E_\nu$ of the excited states and their $X$ and $\mathcal{X}$ amplitudes are obtained by solving (R)STDA eigenvalue problem:
\begin{equation}
\left( \begin{array}{cc}
A_{11} & {A}_{12} \\
{A}_{21} & {A}_{22}
\end{array} \right)
\left( \begin{array}{c}
X^\nu \\
\mathcal{X}^\nu
\end{array} \right)
= E_\nu
\left( \begin{array}{c}
X^\nu \\
\mathcal{X}^\nu
\end{array} \right).
\label{STDAeq}
\end{equation}
with amplitudes $X^\nu = (X^\nu_{ph} ~X^\nu_{\alpha h})^T$ and 
$\mathcal{X}^\nu = (\mathcal{X}^\nu_{pp'hh'} ~\mathcal{X}^\nu_{p\alpha hh' }~\mathcal{X}_{\alpha\alpha' hh'})^T$.  
In the present work we neglect $1$p-$1\alpha$-2h and $2\alpha$-$2$h 
contributions in $Q^\dagger_\nu$ (and $\mathcal{X}^\nu$), therefore we solve the RSTDA 
equations in
p-h $\oplus$ $\alpha$-h $\oplus$ 2p-2h-space.
Furthermore, if we disregard the coupling among 2p-2h states, the matrix $A_{22}$ becomes diagonal, with its elements corresponding to the unperturbed 2p-2h energies:
\begin{equation}
{A}_{22}^{diag}=\delta_{p_1 p_1^{\prime}} \delta_{h_1 h_1^{\prime}} \delta_{p_2 p_2^{\prime}} \delta_{h_2 h_2^{\prime}}\left(e_{p_1}+e_{p_2}-e_{h_1}-e_{h_2}\right).
\label{diagonal}
\end{equation}
We refer to the approximation containing only the diagonal elements of \( A_{22} \), i.e., $A_{22}^{diag}$ as the diagonal RSTDA (RSTDA(d)).
Standard RTDA, i.e., ${A}_{11}$ matrix elements in the RSTDA,  within density functional (DFT) framework are given by (see Ref. \cite{Gam2} or \cite{RingSchuck1980}):
\begin{equation}
\begin{aligned}
\left[A_{\text{11}} \right]_{Ph;{P'}{h'}}&=\frac{1}{2} \sum_{ij}\left.\bigg[\frac{\partial^2 \hat{v}_{ijij}\left[\rho\right]}{\partial\rho_{hP}\partial\rho_{{P'}h'}}\bigg]\right\vert_{\rho~=~\rho^{(0)}}
~+~\sum_j\left.\left[\frac{\partial \hat{v}_{Pjhj}\left[\rho\right]}{\partial \rho_{P'h'}}\right]\right\vert_{\rho~=~\rho^{(0)}} \\
&~~~+~\sum_j\left[\frac{\partial \hat{v}_{{h'}jP'j}\left[\rho\right]}{\partial \rho_{hP}} \bigg]\right\vert_{\rho=\rho^{(0)}}
+~\lambda_{P{P'}}\left[\rho^{(0)}\right] \delta_{h{h'}}\\
&~~~~-~\lambda_{{h'}h}\left[\rho^{(0)}\right] \delta_{P{P'}}~+~\hat{v}_{P{h'}h{P'}}\left[ \rho^{(0)}\right].
\label{dfta11}
\end{aligned}
\end{equation}
The first three terms in eq. \eqref{dfta11} are the so called rearrangement terms, which exist in p-h channels with density dependent couplings (see eq. \eqref{lag}). In the relativistic case, indices $P$ and $P'$ in eq. \eqref{dfta11} are used for both particle and antiparticle states, while $p$, $p_i$ and $p_i'$ stand for particle, and $h$, $h_i$ and $h_i'$ for hole states ($i = 1$ or $ 2$). The present notation for the two-body interaction $\hat{v}_{abcd}$ obeys standard Feynman rules, and $\lambda_{ab}[\rho^{(0)}] = t_{ab} + \sum_k \hat{v}_{akbk}[\rho^{(0)}] + \lambda_{ab}^{\text{rearr}}[\rho^{(0)}]$, which defines the single-particle energies in the relativistic mean-field approximation. For density-independent interactions, the rearrangement term vanishes and only the first two terms remain (see Ref. \cite{Gam2}).
Conversely, when the interaction depends on the density, the rearrangement contribution to the single-particle self-energy ensures consistency with the variational derivation and appears explicitly in both scalar and vector components of the mean field. The explicit form of these rearrangement terms for the DD-PC1 functional is given in \cite{Niksic2008}. The coupling between 1p-1h and 2p-2h configurations is incorporated into the matrix $A_{12}$ element, which is given by the following expression \cite{Providencia}:
\begin{equation}
\begin{aligned}
\left[A_{12}\right]_{{P}h;p_1 p_2 h_1h_2}~=&~ \mathcal{A}(p_1,p_2) \mathcal{A}(h_1,h_2) \delta_{{P}p_1}\delta_{hh_1}\lambda_{h_2 p_2}\left[ \rho^{(0)}\right]
\\&~+~ \mathcal{A}(h_1,h_2) \delta_{hh_1} \hat{v}_{{P} h_2 p_1 p_2} \left[ \rho^{(0)}\right]
\\
&~-~\mathcal{A}(p_1,p_2) \delta_{{P} p_1} \hat{v}_{h_1 h_2 h p_2} \left[ \rho^{(0)}\right],
\end{aligned} 
\label{A12standard}
\end{equation}
with antisymmetrization $\mathcal{A}(a, b) = 1 - P(a, b)$, where $P(a,b)$ is permutation operator which acts on all quantum numbers in $a$ and $b$. For the $A_{12}$ rearrangement term  we have the following term (for the derivation see Ref. \cite{Gam2}):
\begin{align}
\left[A_{12}^{\text{rear. term}}\right]_{Ph;p_1p_2h_1h_2}
&=~\left.\frac{\partial \hat{v}_{h_1h_2p_1 p_2}\left[ \rho\right]}{\partial \rho_{hP}}\right\vert_{\rho=\rho^{(0)}}.
\label{A12rearr}
\end{align}
Upon inclusion of the interaction among 2p–2h configurations in Eq.~\eqref{diagonal}, the complete form of the $A_{22}$ term is obtained as
\begin{equation}
\begin{aligned}
[A^{22}]_{p_1p_2h_1h_2;{p_1^{\prime}}{p_2^{\prime}}{h_1^{\prime}}{h_2^{\prime}}}~&=~\mathcal{A}(p_1,p_2) \mathcal{A}(h_1,h_2) \mathcal{A}({p_1^{\prime}},{p_2^{\prime}}) \delta_{p_1{p_1^{\prime}}}\delta_{h_1{h_1^{\prime}}}\delta_{h_2{h_2^{\prime}}}\lambda_{p_2{p_2^{\prime}}}\\
&~~~-~\mathcal{A}(p_1,p_2)\mathcal{A}(h_1,h_2)\mathcal{A}({h_1^{\prime}},{h_2^{\prime}})\delta_{p_1{p_1^{\prime}}}\delta_{p_2{p_2^{\prime}}}\delta_{h_2{h_2^{\prime}}}\lambda_{{h_1^{\prime}}h_1}\\
~&~~~+~\mathcal{A}(h_1,h_2) \delta_{h_1{h_1^{\prime}}}\delta_{h_2{h_2^{\prime}}}\hat{v}_{p_1p_2{p_1^{\prime}}{p_2^{\prime}}}~+~\mathcal{A}(p_1,p_2)\delta_{p_1{p_1^{\prime}}}\delta_{p_2{p_2^{\prime}}}\hat{v}_{{h_1^{\prime}}{h_2^{\prime}}h_1h_2}\\
~&~~~+~\mathcal{A}(p_1,p_2)\mathcal{A}(h_1,h_2)\mathcal{A}({p_1^{\prime}},{p_2^{\prime}})\mathcal{A}({h_1^{\prime}},{h_2^{\prime}})\delta_{p_1{p_1^{\prime}}}\delta_{h_2{h_2^{\prime}}}\hat{v}_{p_2{h_1^{\prime}}{p_2^{\prime}}h_1}.
\end{aligned}
\end{equation}
Applying the same procedure as for $A_{12}^{\text{rear}}$ terms, the rearrangement contribution to $A_{22}$ is given by:
\begin{align}
\left[A_{22}^{\text{rear}}\right]_{p_1 p_2 ij;p_1^\prime p_2^\prime kl}~&=~\left.\frac{1}{2}\sum_{hh'}\mathcal{A}(p_1,p_2) \mathcal{A}(p_1^\prime,p_2^\prime)\frac{\hat{v}_{hh'hh'}\left[\rho\right]}{\rho_{p_2^\prime p_2}}\right\vert_{\rho=\rho^{(0)}}\delta_{p_1 p_1^\prime}\delta_{ik}\delta_{jl}\nonumber\\
&~-~\left.\frac{1}{2}\sum_{hh'}\mathcal{A}(i,j) \mathcal{A}(k,l)\frac{\hat{v}_{hh'hh'}\left[\rho\right]}{\rho_{ik}}\right\vert_{\rho=\rho^{(0)}}\delta_{p_1 p_1^\prime}\delta_{p_2 p_2^\prime}\delta_{jl}.
\end{align}
Both $A_{12}^{\text{rear}}$ and $A_{22}^{\text{rear}}$ terms should contain all p-h channels as in the relativistic Hartree ground state in order to have self-consistent RSTDA and RSSTDA calculations. On the other hand, subtraction procedure from Ref. \cite{Gambacurta2015} is modified by the reduction from SRPA to the STDA case, i.e. ignoring terms which contain $B$ matrices:
\begin{equation}
A_{11^{\prime}}^S=A_{11^{\prime}}+\sum_{2 2^\prime} A_{12}\left(A_{22^\prime}\right)^{-1} A_{2^\prime 1^{\prime}}.
\end{equation}
In order to obtain subtracted RSTDA equations, one needs to replace $A_{11}$ from Eq. (\ref{STDAeq}) with $A_{11}^S$. The general coupled form of $A_{11}^{J}$, $A_{12}^J$ and $A_{22}^J$ matrices may be found in Ref. \cite{Papakonstantinou2010}. The mathematical procedure for evaluating both uncoupled and coupled relativistic two-body matrix elements for selected p-h channels is presented in Ref.~\cite{Vale2}.

One of the quantities which we want to investigate is transition strength distribution defined as:
\begin{equation}
\begin{aligned}
R(E)~&=~\sum_\nu |\langle \nu | \mathbf{\hat{F}} | 0 \rangle|^2\delta(E - E_\nu)\\
&=~\sum_\nu B_F(E_\nu)\delta(E - E_\nu).
\end{aligned}
\end{equation}
In this work we only consider operators of spin independent modes given by expressions:
\begin{equation}
\begin{aligned}
\hat{F}_{\lambda M}^{\text{IS}} ~=&~ \sum_i r_i^k Y_{\lambda M} (\Omega_i),   \\
\hat{F}_{\lambda M}^{\text{IV}} ~=&~ \sum_i r_i^k Y_{\lambda M} (\Omega_i) \hat{t}_z(i),
\end{aligned}    
\end{equation}
for the isoscalar (IS) and isovector (IV) channel, respectively. For the $\lambda = 0$, i.e. the monopole case, $k = 2$, while for other cases $k = \lambda$. The reduced transition strength $B_F(E_\nu)$ is calculated only from 1p-1h configuration, i.e. obtained from $X_{ph}$ amplitudes and single-particle transition matrix elements $f_{ph} = \langle p | \hat{F} | h \rangle$ and $f_{\alpha h} = \langle \alpha | \hat{F} | h \rangle$:
\begin{equation}
B_F^{\lambda}(E_\nu) ~=~   |\sum_{p h}  f_{ph}^{\lambda (\nu)} X_{p h }^{\lambda (\nu)} + \sum_{\alpha h}  f_{\alpha h}^{\lambda (\nu)} X_{\alpha h }^{\lambda (\nu)}|^2. 
\end{equation}
The moments of transition strength weighted by energy are defined as:
\begin{equation}
m_k~=~\sum_{\nu} E_\nu^{k} B_F^{\lambda}(E_\nu),
\end{equation} 
from which the centroid energy is obtained as
\begin{equation}
\bar{E} ~=~ \frac {m_1}{m_0},
\end{equation}
where the moments $m_k$ may be evaluated over the whole excitation spectrum (positive, negative or both parts), in the energy
region of a resonance or in the experimental window. The width of a distribution or a
resonance can be expressed as
\begin{equation}
\Delta ~=~ \sqrt{\frac{m_2}{m_0}-\bar{E}^2}.
\end{equation}
If we ignore 1p-1$\alpha$-2h and $2\alpha$-2h configurations the RSTDA normalization condition is given by expression: 
\begin{equation}
\sum_{ph}|X_{ph}^\nu|^2 + \sum_{\alpha h} |X_{\alpha h}^\nu|^2 + \sum_{p_1 p_2 h_1 h_2} |X_{p_1 p_2 h_1 h_2}^\nu|^2 ~=~N_1 + N_2 ~=~ 1.
\label{norma}
\end{equation}
For the 1p-1h, 1$\alpha$-1h and 2p-2h configuration, the contribution of the specific configuration to the RSTDA norm is
\begin{equation}
\begin{aligned}
N_{1p1h} ~=&~ \sum_{ph}|X_{ph}^\nu|^2,\\
N_{1\alpha 1h} ~=&~ \sum_{\alpha h}|X_{\alpha h}^\nu|^2,\\
N_{2p2h} ~=&~ \sum_{p_1 p_2 h_1 h_2}|X_{p_1 p_2 h_1 h_2}^\nu|^2.    
\end{aligned}    
\end{equation}
with $N_1 = N_{1p1h} + N_{1\alpha1h}$ and $N_2 = N_{2p2h}$. In the RTDA case, the normalization condition consist of the first two terms in eq. \eqref{norma}.

\section{Results and discussion}\label{four}
In the following we present the results on the ISGMR, IVGMR, ISGQR and IVGQR properties calculated within several variants of the relativistic STDA. The ground state single-particle energies, and upper and lower components of spinors are obtained in the Relativistic-Hartree approximation by solving the Dirac equations using a basis of spherical harmonic oscillator wave functions \cite{Nik}. The basis includes states up to the maximum principal quantum number $N = 20$ in the case of GMR and $N = 10$ in the case of GQR, both for ground and excited states. While this choice for the later is primarily motivated by computational limitations, it may contribute to the slight overestimation of excitation energies observed in the RSTDA results. The selection of 1p-1h and 2p-2h configurations follows the rules of isospin, parity, and total angular momentum. An energy cutoff is also applied, i.e. for 1p-1h {configurations}, unperturbed transition energies are considered up to 200 MeV, while for 1$\alpha$-1h configurations, involving transitions to empty negative-energy Dirac sea states denoted by $\alpha$, are taken into account up to 2000 MeV. In the case of 2p-2h configurations, the appropriate energy cutoff is determined to ensure convergence in the RSSTDA calculations.  

\subsection{Transition strength distributions of ISGMR and IVGMR}

\begin{figure}[h]
  \includegraphics[width=0.9\linewidth]{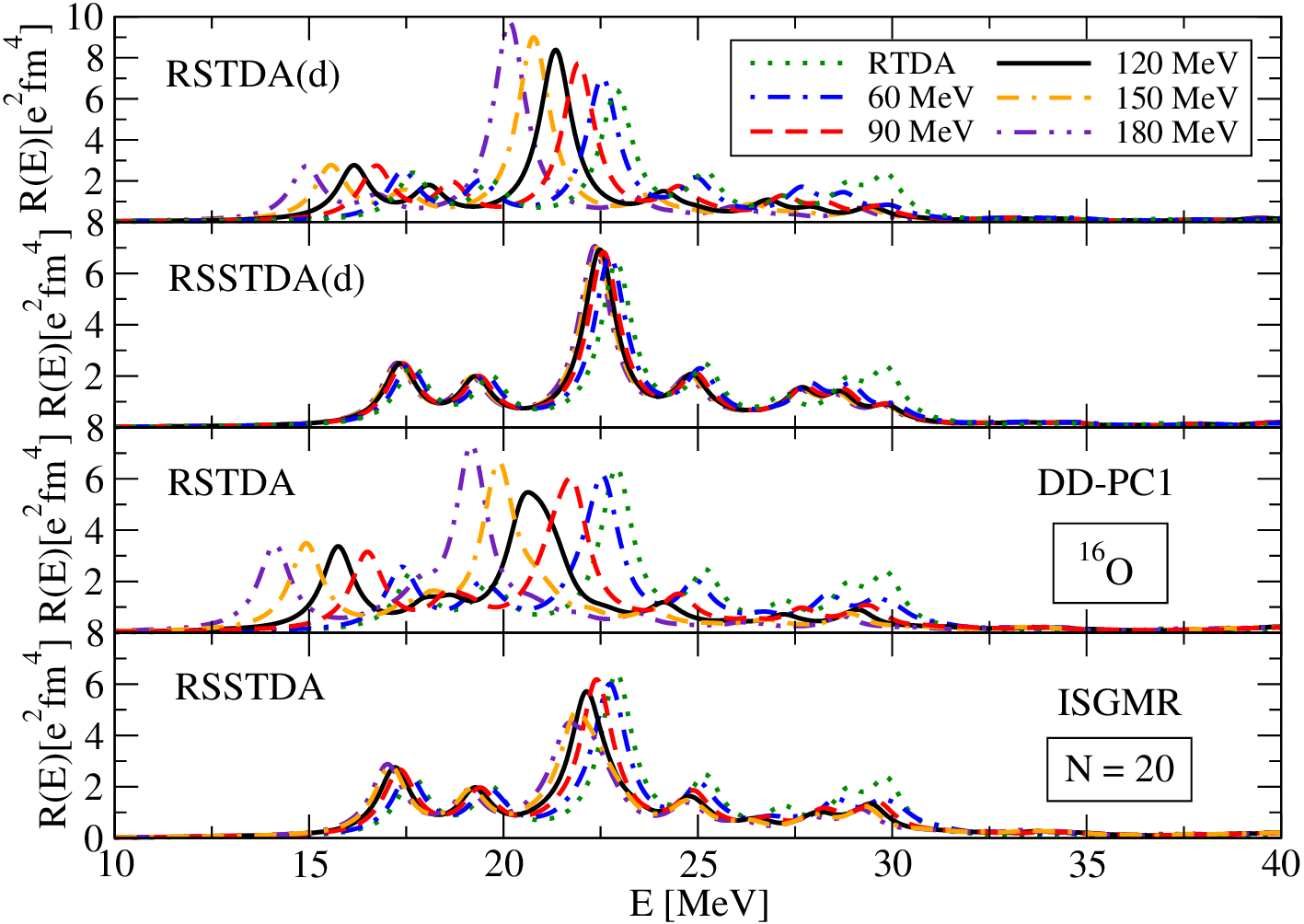}
  \caption{ISGMR strength distributions for $^{16}$O using TDA, RSTDA and RSSTDA with the full $A_{22}$ form and diagonal approximation (denoted with (d)) as functions of excitation energy, for different 2p-2h energy cutoffs as given in the legend. For details see the text.}
  \label{fig:ISGMR:N20}
\end{figure}

Figure \ref{fig:ISGMR:N20} presents the ISGMR (\(0^+\)) strength distribution of $^{16}$O calculated using the relativistic point-coupling model with the DD-PC1 parameterization and principal quantum number $N = 20$. To ensure a smooth representation, all discrete results are folded with a Lorentzian function of 1 MeV width. The calculations include both standard and subtracted form of RSTDA in its full and diagonal approximation. For a comparison, the results of the RTDA are also shown in Fig. \ref{fig:ISGMR:N20}. {In} the full form of RSTDA and RSSTDA, all rearrangement terms are included in submatrices $A_{11}$, $A_{12}$ and $A_{22}$, while the diagonal case does not include rearrangement terms in the $A_{22}$ submatrix. The RSTDA and diagonal case RSTDA(d) calculations are shown for 2p-2h energy cutoffs of 60, 90, 120, 150 and 180 MeV, experiencing progressively larger shifts toward lower excitation energies with increasing cutoff energy. In the calculations with subtraction,  RSSTDA and RSSTDA(d) results show convergence of all major peaks $B(E_i) \gtrsim 0.1$ e$^2$fm$^4$ within the cutoff change from 150 to 180 MeV, with {peak position shifts} towards lower energies by $\Delta E_x^{SSTDA} \lesssim 150$ keV and $\Delta E_x^{SSTDA(d)} \lesssim 100$ keV, respectively. As previously noted in Refs.~\cite{Yang2021,Gambacurta2015}, the use of the diagonal approximation induces only minor changes in the excitation spectrum within the subtracted framework. This behavior is confirmed in our relativistic implementation as well, where slightly more pronounced downward energy shifts of a few hundred keV in RSSTDA are observed when comparing results with same 2p-2h cutoff energies. Furthermore, the RSSTDA(d) in this work is calculated self-consistently at each step, whereas in Ref.~\cite{Gambacurta2015}, the full $A_{22}$ matrix is used throughout, except during the subtraction step, where only the diagonal part of $A_{22}$ is employed in the inversion procedure. Therefore, it is important to emphasize that the definition of diagonal RSSTDA adopted here does not fully coincide with that in previous works. As discussed later in the text, this difference has a non-negligible impact on the preservation of the inverse energy-weighted sum rule ($m_{-1}$).

\begin{figure}[h]
  \includegraphics[width=0.9\linewidth]{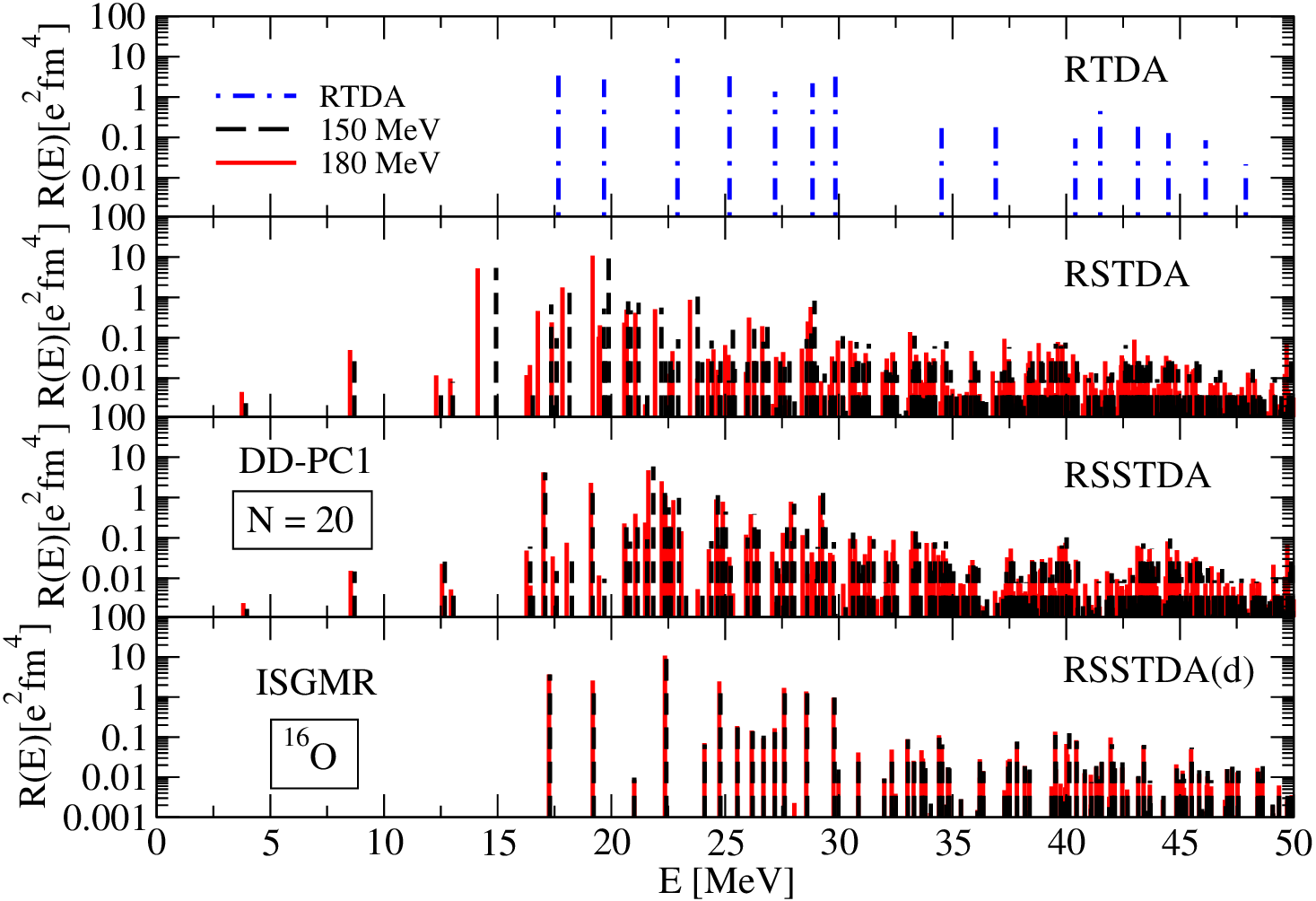}
  \caption{{Discrete ISGMR strength distribution in $^{16}$O for RTDA, RSTDA, RSSTDA and RSSTDA(d)  as functions of excitation energies for 2p-2h energy cutoffs 150 and 180 MeV.}}
  \label{fig:ISGMR:N20:Discrete}
\end{figure}

Figure \ref{fig:ISGMR:N20:Discrete} shows the discrete strength distribution for the ISGMR transition, revealing the fragmentation of states in more details, with a large number of emerging states predominantly of the 2p-2h nature ($N_2 \gtrsim 0.99$ from Eq. (\ref{norma})).
When compared to RTDA, for $E_{\text{cutoff}} = 150$ MeV all major peaks ($B \gtrsim 0.1 $ e$^2$fm$^4$) predominantly 1p-1h $(N_1 \gtrsim 0.8)$ and the ones with mixed nature below $E_x = 30$ MeV in RSTDA and RSTDA(d) are shifted toward lower energies by up to approximately 3 MeV and 2.5 MeV, respectively. With further increase of the cutoff energies, the standard RSTDA does not provide convergence of the excitation energies. For RSSTDA and RSSTDA(d), the energy shifts are limited to up to $\approx1$ MeV when compared to RTDA. The key difference between RSSTDA and RSSTDA(d) lies in the absence of interaction between 2p-2h states in the latter, which results in somewhat less pronounced fragmentation. This, in turn, implies a smaller number of predominantly 2p-2h states with reduced transition strengths ($B \gtrsim 0.001$ e$^2$fm$^4$) between the main predominantly 1p-1h and mixed peaks in the lower and intermediate part of excitation spectrum ($E_x\lesssim40$ MeV). 

The calculated \(0^+\) state at approximately 4~MeV may be associated with the experimentally observed \(0^+\) state at 6.05~MeV \cite{A16}, which is commonly interpreted as an \(\alpha\)-cluster resonance. The subsequent calculated \(0^+\) states appear at 8.7, 12.7, and 13.1~MeV, with reduced transition strengths of the order of \(10^{-2}\)e$^2$fm$^4$. On the experimental side, in addition to the well-known state at 6.05~MeV, other \(0^+\) candidates have been observed around 12.05 and 14.01~MeV, which may be related to fragmented components of the ISGMR \cite{A16}. The energy discrepancy between RSSTDA and experimental values can be attributed to the choice of the energy density functional, which was originally optimized for medium to heavy nuclei and may not provide optimal reproduction of absolute excitation energies in light nuclei such as \(^{16}\)O. Nevertheless, the structure of the first few states is overwhelmingly of 2p-2h nature, with a total 2p-2h norm \(N_2\) exceeding \(99\%\).

The results for the \(0^+\) states below 16~MeV indicate that, in the absence of rearrangement terms and a fully self-consistent formulation, the RSSTDA approach is unable to describe even qualitatively the low-energy part of the monopole excitation spectrum in \(^{16}\)O. In particular, the subtracted version within the diagonal approximation to the \(A_{22}\) matrix, denoted as RSSTDA(d), does not succeed in capturing the experimental low-energy structure. This highlights the essential role of dynamical correlations and a full treatment of the 2p-2h subspace in achieving a realistic description of \(0^+\) excitations. 

The moments of the transition strength weighted by energy $m_k$, centroid energy $\bar{E}$, width $\Delta$ of the distribution, and the percentage of the energy weighted sum rule (EWSR) for the ISGMR, along with experimental values from \cite{Lui2001}, are shown in Tab.~\ref{tabIS}. The total strength of the positive part of the ISGMR excitation spectrum, represented by the $m_0$ moment, varies only marginally between RTDA, RSSTDA(d), and RSSTDA with 2p-2h cutoff beyond $E_{\text{cutoff}} = 150$ MeV. The difference between RTDA and RSSTDA(d) is minimal, $| m_0^{\text{RTDA}}  - m_0^{\text{RSSTDA(d)}} | \approx 8.0\times10^{-3}~\text{e}^2\text{fm}^4$, while the difference between RSSTDA(d) and RSSTDA is somewhat larger, around $9.1\times10^{-3}~\text{e}^2\text{fm}^4$. The $m_1$ moment, which reflects the total energy-weighted strength, increases significantly from RTDA to RSSTDA(d) and RSSTDA, reaching $m_1^{\text{RSSTDA}} = 774.098~\text{e}^2\text{MeVfm}^4$, compared to $691.180~\text{e}^2\text{MeVfm}^4$ for RTDA, which is a 12.0\% increase when compared to RTDA. In Ref.~\cite{Yang2021}, an increase of approximately 7.5\% (9.5\%) in the  \( m_1 \) momentum was obtained in SRPA(d) (SRPA) compared to the RPA value.
However, when limited to the experimental window \cite{Lui2001}, a significant reduction in $m_1$ is observed across all models, i.e., $m_1$ drops to $629.141$ in RTDA, $583.897$ in RSSTDA(d), and $556.526~\text{e}^2\text{MeVfm}^4$ in RSSTDA, respectively. These results indicate that a considerable portion of the transition strength in RSSTDA(d) and RSSTDA is redistributed outside the experimentally accessible energy range.

\begin{table}[h]
\begin{center}
    \caption{Energy moments and energy weighted sum rule (EWSR) of isoscalar $0^+$  transition obtained by the RTDA, RSSTDA(d) and RSSTDA calculations for $^{16}$O with DD-PC1 interaction. For RSSTDA(d) and RSSTDA calculations the energy 2p-2h cutoff is taken $E_{cutoff}=150$ MeV. The values in round brackets are calculated within experimental window from \cite{Lui2001}.}
    \begin{tabular}{c c c c c}
        \hline\hline
         $ $ & RTDA & RSSTDA(d) & RSSTDA & Exp. \cite{Lui2001}\\
        \hline
        \multicolumn{5}{c}{$N = 20$} \\
        $m_0$ [e$^2$fm$^4$] & 27.586 (26.343) & 27.578 (25.225) & 27.578 (24.316) &   \\
        $m_1$ [e$^2$MeVfm$^4$]  & 691.180 (629.141) & 758.322 (583.897)  & 774.098 (556.526) &  \\
        $m_2$ [e$^2$MeV$^2$fm$^4$] & 19000.50 (15455.12) & 29537.18 (13966.58) & 30665.05 (13285.55) & \\
        $m_{-1}$ [e$^2$MeV$^{-1}$fm$^4$] & 1.162 (1.135) & 1.162 (1.124) & 1.162 (1.104) & \\
        $m_1/m_0$[MeV] & 25.055 (23.883) & 27.497 (23.148) & 28.069 (22.887) & 21.13$\pm$0.49\\      
        $\sqrt{m_1/m_{-1}}$ [MeV] & 24.391 (23.544) & 25.548 (22.793) & 25.813 (22.453)  & 19.63$\pm$0.38\\
        $\sqrt{m_3/m_{1}}$ [MeV] & 30.679 (24.903) & 51.026 (24.341) & 50.798 (24.424)  & 24.89$\pm$0.59\\
        $\Delta$ [MeV] & 7.811 (4.037) & 17.747 (4.227) & 18.001 (4.749) & 8.76$\pm$1.82 \\
        EWSR[\%] & 95.74 (87.15) & 105.04 (80.88) & 107.22 (77.09) & 48$\pm$10 \\
        \hline\hline
    \end{tabular}
    \label{tabIS}
\end{center}
\end{table}

The centroid energy, given by $m_1/m_0$, provides insight into the average excitation energy of the transition strength distribution. For the full spectrum, this value increases across the models: $25.055$ MeV (RTDA), $27.497$ MeV (RSSTDA(d)), and $28.069$ MeV (RSSTDA). A shift of several MeV toward higher energies in the SSRPA distribution of monopole transition strength, relative to the RPA case, has already been observed upon increasing the 2p–2h cutoff up to 70 MeV in Ref.~\cite{Gambacurta2015}. When restricted to the experimental window (11-40 MeV), the centroid energies are systematically reduced in all models, i.e., by $1.17$ MeV in RTDA, $4.35$ MeV in RSSTDA(d), and $5.18$ MeV in RSSTDA, yielding values of $23.883$ MeV, $23.148$ MeV, and $22.887$ MeV, respectively. Compared to the experimental value of $21.13 \pm 0.49$ MeV, all theoretical models slightly overestimate the centroid energy, with RSSTDA showing the smallest discrepancy in the experimental window ($|\Delta E| \approx 1.76$ MeV).

The effective transition energy, defined as $\sqrt{m_1/m_{-1}}$, increases from $24.391$ MeV in the RTDA to $25.548$ MeV in the RSSTDA(d), and further to $25.813$ MeV in the RSSTDA. This corresponds to a relative increase of $4.8\%$ and $5.8\%$ with respect to the RTDA, respectively. All theoretical values overestimate the experimental result of $19.63 \pm 0.38$ MeV, which suggests that additional effects, such as more complex configurations than 2p-2h, might be required to reduce the effective transition energy. In Ref.~\cite{Yang2021}, the Skyrme SRPA (SRPA(d)) calculations resulted in $m_1 = 724.324$ ($737.777$) e$^2$MeVfm$^4$ and $m_{-1} = 1.169$ ($1.147$) e$^2$MeV$^{-1}$fm$^4$, which leads to values $\sqrt{m_1/m_{-1}} = 24.892$ ($25.362$) MeV. The discrepancy of the $m_{-1}$ moment between SSTDA and SSTDA(d), reported in Ref. \cite{Yang2021}, was not observed in the present work. This is attributed to the fact that the subtraction procedure was implemented in a self-consistent manner within the relativistic extension of the STDA framework, specifically adapted for this formalism.

The ratio $\sqrt{m_3/m_{1}}$, which emphasizes the contribution of high-energy components in the transition strength distribution, significantly increases in RSSTDA(d) and RSSTDA compared to RTDA. The RTDA predicts $30.679$ MeV, whereas RSSTDA(d) gives $51.026$ MeV (an increase of $66.3\%$), and RSSTDA predicts $50.798$ MeV (an increase of $65.6\%$). This suggests that correlations in RSSTDA(d) and RSSTDA strongly redistribute strength towards higher excitation energies. However, when considering the energy range $11 - 40$ MeV, the RTDA, RSSTDA(d), and RSSTDA results are close to the experimental value of $(24.89 \pm 0.59)$ MeV \cite{Lui2001}, but when comparing RSSTDA(d) and RSSTDA, RSSTDA is showing the smallest discrepancy ($|\sqrt{m_{3}^{\text{exp}}/m_{1}^{\text{exp}}}-\sqrt{m_{3}^{\text{RSSTDA}}/m_{1}^{\text{RSSTDA}}}|\approx 0.47$ MeV).

The EWSR saturation is nearly complete in all theoretical models, with $95.74\%$ for RTDA, $105.04\%$ for RSSTDA(d), and $107.22\%$ for RSSTDA. Refs.~\cite{Gambacurta2015, Yang2021} also report values that are either close to or slightly exceed 100\% exhaustion of the energy-weighted sum rule (2p-2h $E_{cutoff}\lesssim70$ MeV), with SSRPA(d) yielding marginally higher strength than SSRPA.  In the present work, the trend is reversed, likely due to the self-consistent formulation of the subtraction method within the relativistic STDA. However, within the experimental window, these percentages drop significantly ($87.15\%$ for RTDA, $80.88\%$ for RSSTDA(d), and $77.09\%$ for RSSTDA), whereas the experiment reports only $(48 \pm 10)\%$. This indicates that a significant portion of the theoretically predicted strength lies outside the experimental window from \cite{Lui2001}, particularly in RSSTDA(d) and RSSTDA calculations, and further experimental studies are required to reduce these discrepanices. Although not explicitly shown in Table~\ref{tabIS}, both RSTDA and RSTDA(d) results for the ISGMR appear to preserve the \(m_0\) and \( m_1 \) moment, and thus the EWSR, when compared to the corresponding RTDA values. In contrast, the SRPA calculations in Ref.~\cite{Papakonstantinou2010} exhibit a slight deviation in the \( m_0 \) moment, amounting to approximately 3\% compared to the corresponding RPA result, while preserving $m_1$ moment.

\begin{figure}[h]
  \includegraphics[width=0.9\linewidth]{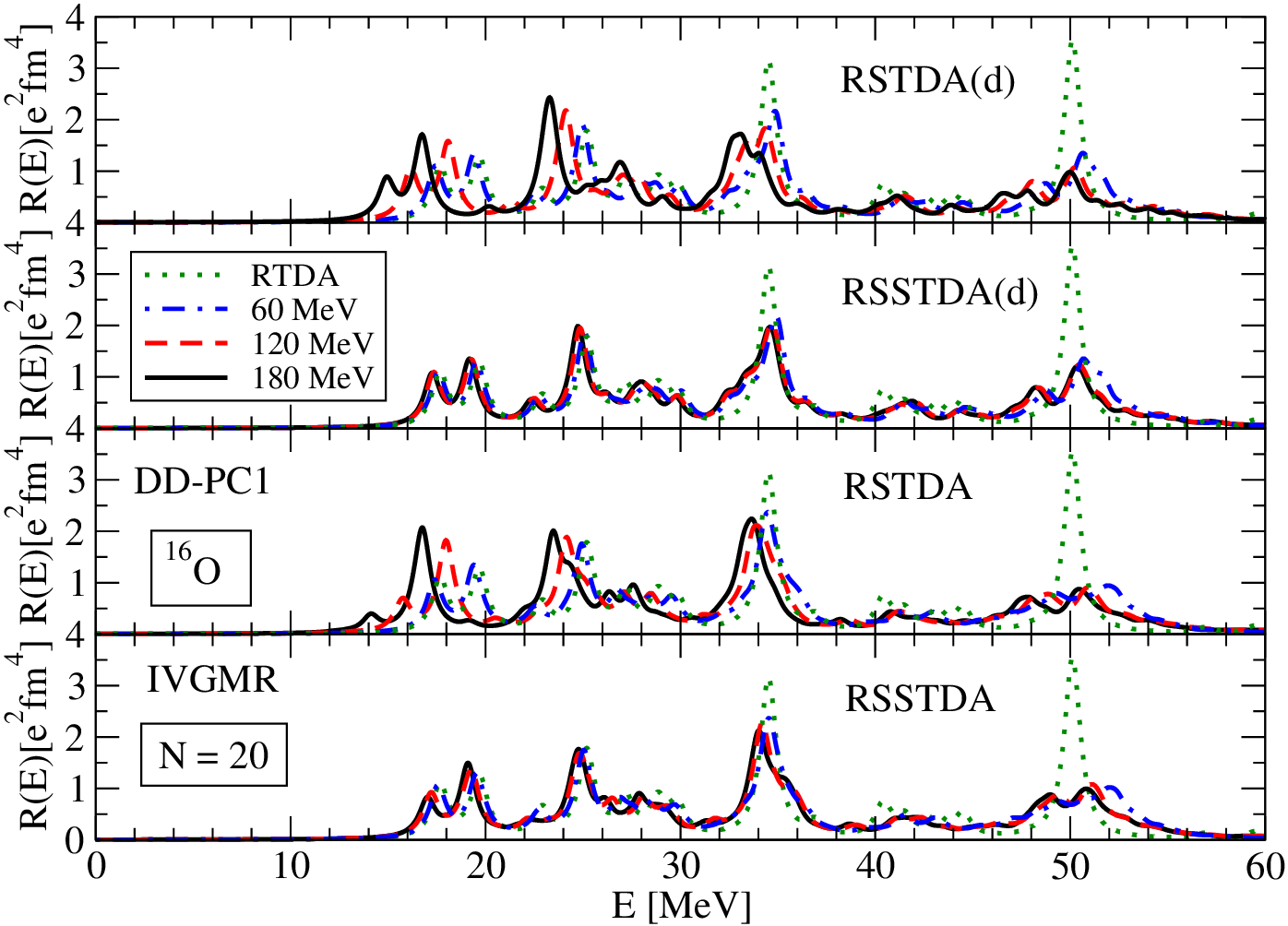}
  \caption{The same as Fig. \ref{fig:ISGMR:N20}, but for IVGMR strength distribution.}
  \label{fig:IVGMR:N20}
\end{figure}

\begin{figure}[h]
  \includegraphics[width=0.9\linewidth]{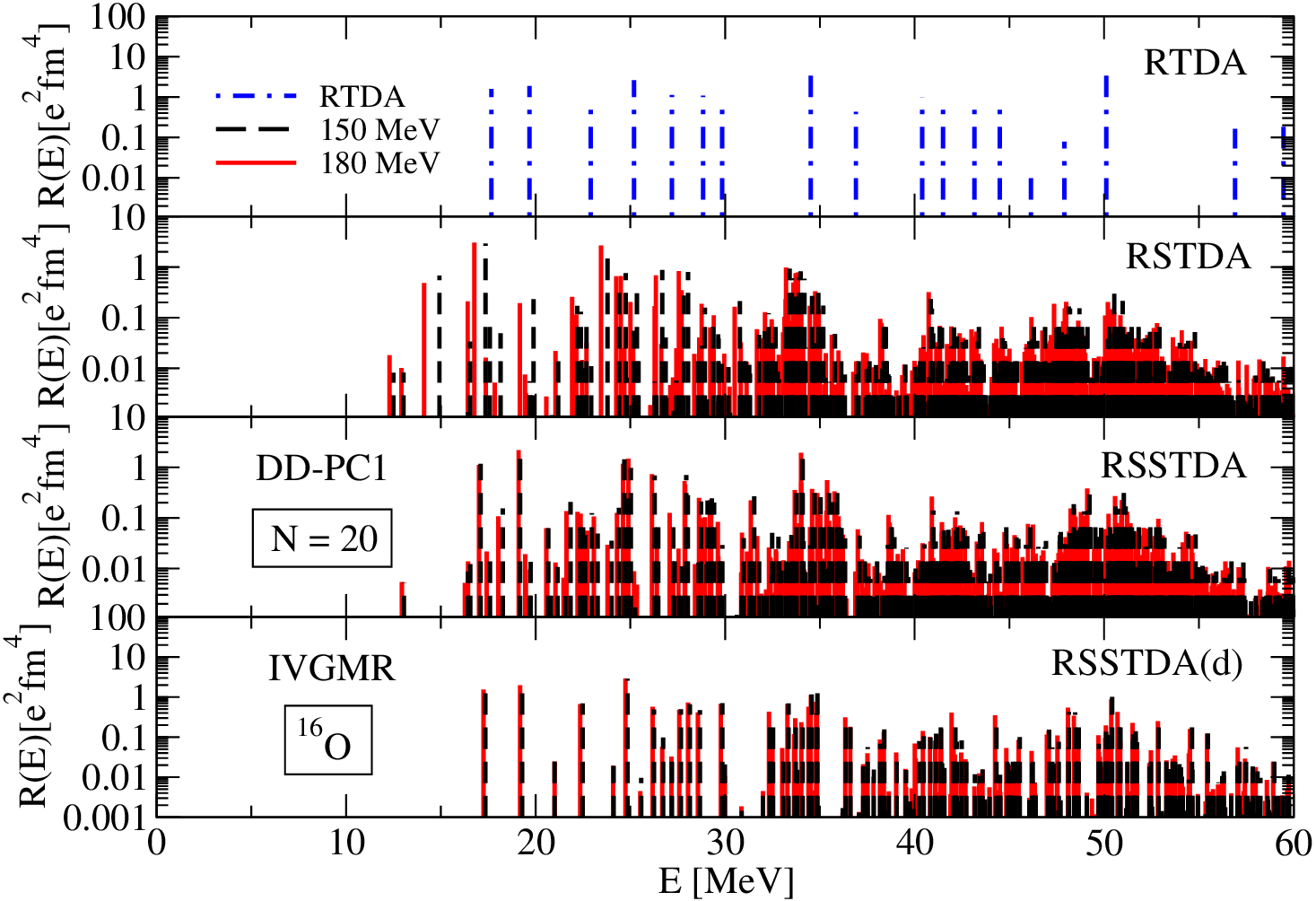}
  \caption{The same as Fig. \ref{fig:ISGMR:N20:Discrete}, but for IVGMR strength distribution.}
  \label{fig:IVGMR:N20:Discrete}
\end{figure}

In the following we investigate the properties of another monopole type of excitation, IVGMR.
Figure~\ref{fig:IVGMR:N20} presents the IVGMR transition strength distributions for \(^{16}\)O, calculated using previously discussed approximations (RTDA,  {RSTDA(d)},  RSSTDA(d), RSTDA, and RSSTDA)  within the relativistic point-coupling model with DD-PC1 parametrization and for varying energy cutoffs for 2p-2h configurations, (60, 120, and 180~MeV). While Fig.~\ref{fig:IVGMR:N20} displays Lorentzian-smoothed distributions, Fig.~\ref{fig:IVGMR:N20:Discrete} shows the corresponding discrete strength distributions for RTDA, RSTDA, RSSTDA, and RSSTDA(d) and cutoff energies 150 and 180 MeV. The RTDA approximation shows stronger peaks at higher energies of the spectrum. In contrast, the RSTDA approximation, which includes rearrangement terms, reduces the strength of high-energy peaks and causes the major peaks to shift towards lower energies, with a noticeable fragmentation of states, particularly as the energy cutoff increases. This fragmentation indicates a more complex interaction among the nuclear states. RSSTDA and RSSTDA(d), which include further refinements, show less fragmentation and a more stable distribution. RSSTDA(d), in particular, exhibits fewer 2p-2h states due to the absence of interactions between these configurations, resulting in somewhat reduced transition strengths and a more concentrated distribution in the major predominantly 1p-1h states. While the energy cutoff of 150 MeV causes a significant shift in the peak positions, up to $\approx2.5$ MeV in RSTDA when compared to RTDA, RSSTDA and RSSTDA(d) show more stability, with shifts generally confined to $\approx1$ MeV. As the cutoff energy reaches 180 MeV, the RSTDA peaks display an enhanced shift toward lower energies by $\Delta E_x\approx 1$ MeV for states with $E_x\lesssim35.0$ MeV, while RSSTDA and RSSTDA(d) demonstrate reasonable convergence, with peaks additionally shifting towards lower energies only by approximately 150 keV and 100 keV, respectively. 

\subsection{Transition strength distributions of ISGQR and IVGQR}

\begin{figure}
  \includegraphics[width=0.9\linewidth]{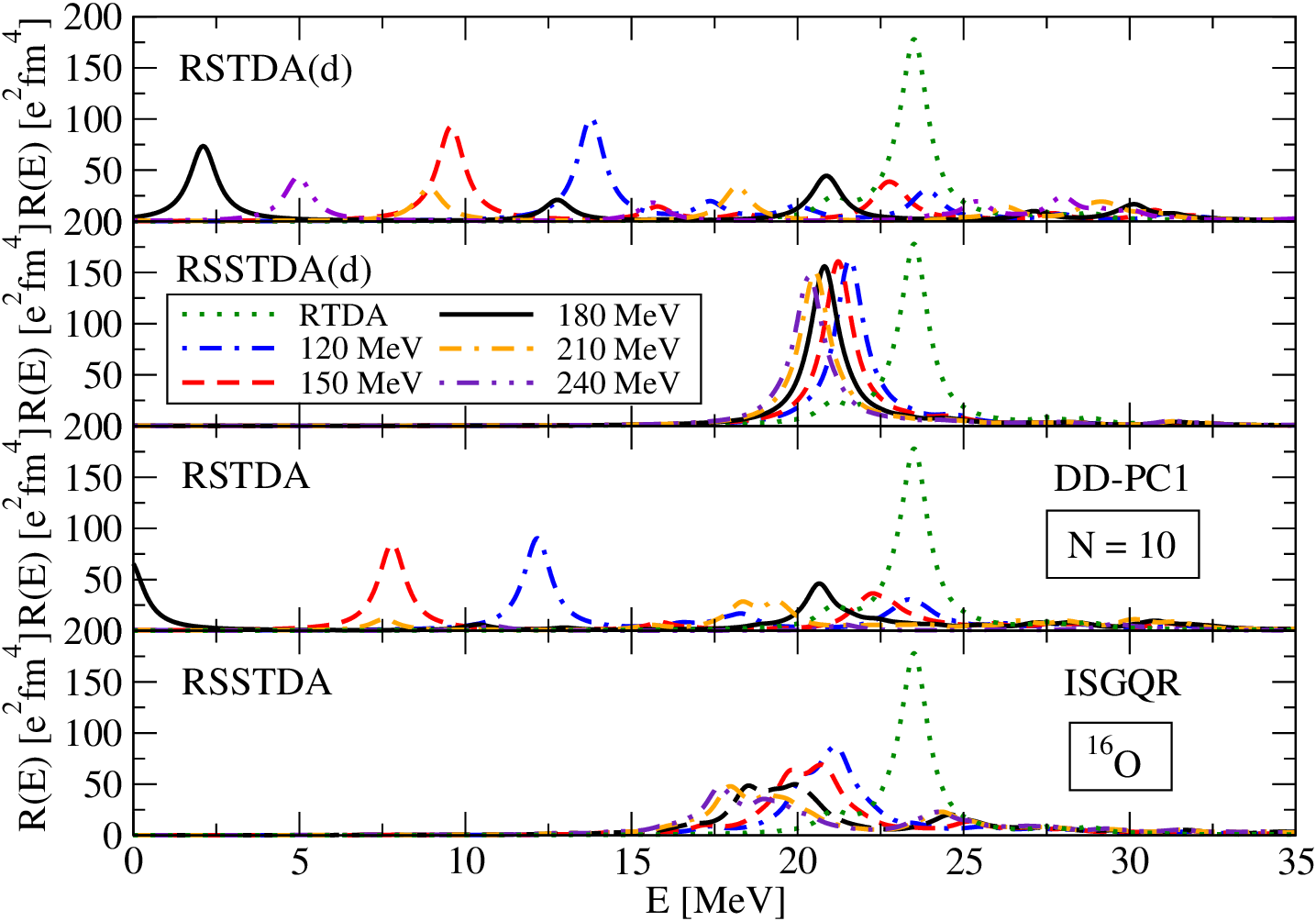}
  \caption{{The same as Fig. \ref{fig:ISGMR:N20}, but for ISGQR strength distribution. }}
  \label{fig:ISGQR:N14}
\end{figure}

\begin{figure}
  \includegraphics[width=0.9\linewidth]{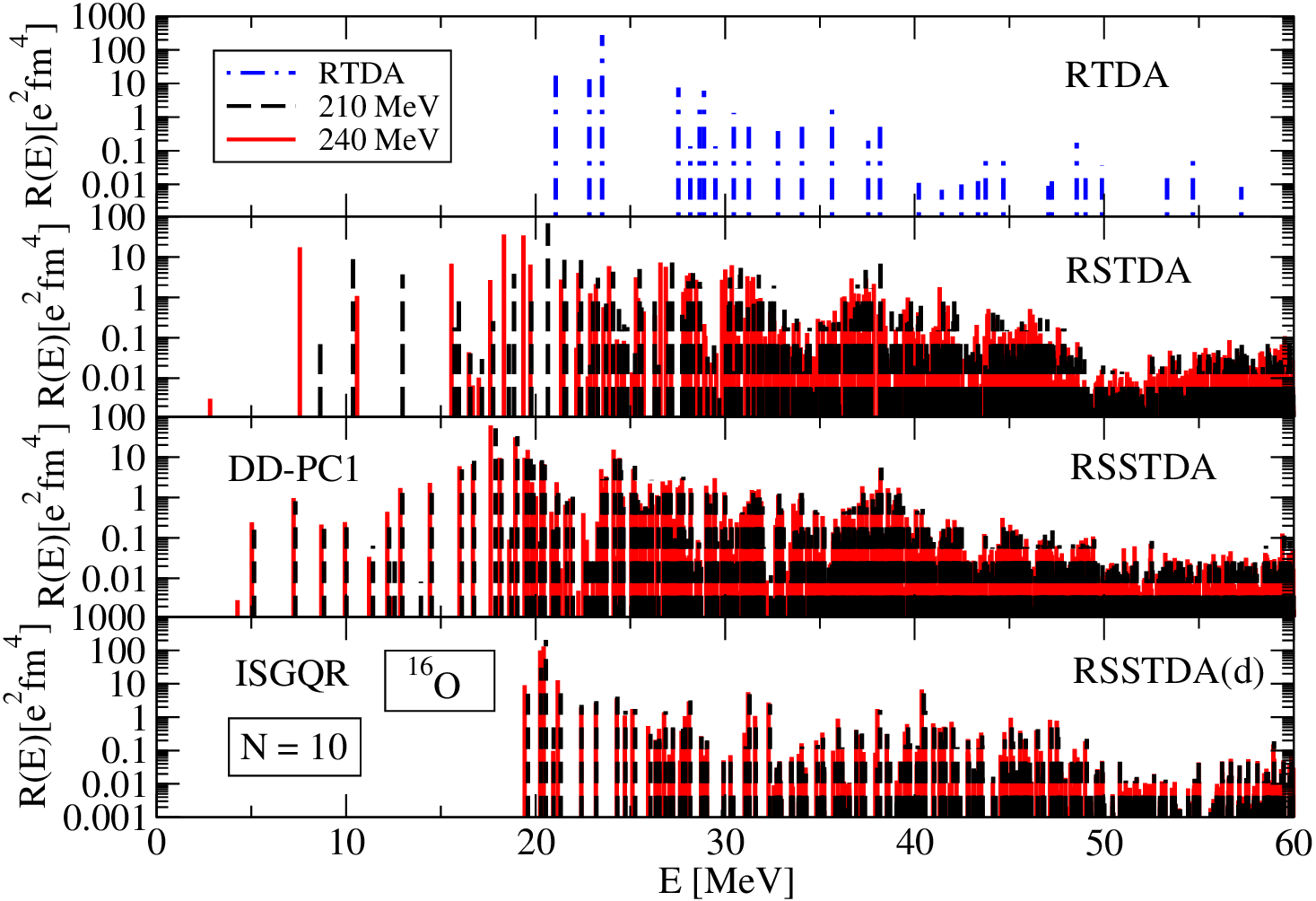}
  \caption{The same as Fig. \ref{fig:ISGMR:N20:Discrete}, but for ISGQR strength distribution. }
  \label{fig:ISGQR:N14:discrete}
\end{figure}

The strength distributions of the isoscalar quadrupole giant resonances (ISGQR) in $^{16}$O, calculated using the RSTDA, RSTDA(d), RSSTDA, and RSSTDA(d) approximations for various 2p-2h energy cutoffs (from $120$  to $240$ MeV), are shown in Fig. \ref{fig:ISGQR:N14}. The corresponding transition strengths for discrete spectra obtained with 2p-2h energy cutoffs $210$ and $240$ MeV are presented in Fig. \ref{fig:ISGQR:N14:discrete}. Due to computational limitations, here the basis used in calculations includes states up to the maximum principal quantum number $N = 10$.
As shown in Fig. \ref{fig:ISGQR:N14}, the dominant peak in RTDA splits into multiple peaks in RSTDA upon the inclusion of interactions between 1p-1h and 2p-2h configurations, even for lower 2p-2h cutoff values such as 60 MeV.
The RSTDA approximation, shows significant fragmentation of the strength distribution, with peaks shifting towards lower energies much faster than ISGMR or IVGMR as the cutoff increases. 
In contrast to the ISGMR and IVGMR transitions, where only small difference between RSTDA and RSTDA(d) spectrum is observed, the interaction between 2p-2h configurations plays crucial role in shaping the ISGQR transition strength distribution, causing both larger shifts of the eigenvalues to the lower excitation energies and changes of the transition strengths. For the lowest dominant peak (below 15 MeV) the difference between RSTDA and RSTDA(d) eigenvalue is $\Delta E_x ~\approx 2.5$ MeV ($\Delta E_x ~\approx 2$ MeV) and similarly for other peaks for the energy cutoff $E_{cutoff}~=~150$ MeV ($E_{cutoff}~=~120$ MeV). For higher energy 2p-2h cutoffs significant part of the transition strength falls into the negative excitation energy region. A similar issue was also reported in the context of Gamow–Teller transitions by a recent nonrelativistic STDA study \cite{Minato2016}, where a significant portion of the strength was found in the low-energy region, within range [–20, 10] MeV. The accumulation of strength in this region is directly related to the infrared instability of the STDA.

The RSSTDA exhibits more stable distributions with less pronounced fragmentation compared to RSTDA, especially as the energy cutoff increases. The subtraction method eliminates infrared divergence and further fragments the dominant transition into multiple peaks distributed between $\approx$4 and 25 MeV. 
RSSTDA is characterized by 12 low lying 2p-2h states ($N_2 \gtrsim 99\%$) for  $E_{cutoff} \gtrsim 60$ MeV, which shape excitation spectrum bellow $E_x\lesssim 17$ MeV and show weak energy dependence on the choice of the 2p-2h cutoff. For cutoff energies $E_{cutoff} \lesssim150$ MeV, predominantly 1p-1h and mixed peaks below 25 MeV in RSSTDA shift toward lower excitation energies up to a few hundred keV for every $30$ MeV. However, beyond $E_{\text{cutoff}} = 150$ MeV, the transition strength distribution exhibits a tendency toward reasonable convergence, i.e., energy shifts which tend to stabilize below $\approx100$ keV per $\Delta E_{\text{cutoff}} = 30$ MeV when the cutoff is increased from 210 to 240 MeV.
A similar behavior has been observed in non-relativistic SRPA calculations based on the Skyrme energy density functional in Ref. \cite{Gambacurta2015}. On the other hand, most of the strength in RSSTDA(d) is contained in one state, which indicates significantly less pronounced fragmentation, while peaks shift toward lower energies at similar convergence rate. For example, energy shift of the most contributing state decreases by $\approx30-40\%$ every 30 MeV in the energy range of cutoffs from 90 MeV to 240 MeV.

To obtain a satisfactory convergence of the excitation spectrum below 25 MeV, a sufficiently large 2p–2h configuration space must be included.  For the present DD‑PC1 calculations in $^{16}$O this requires extending the 2p–2h energy cut‑off to at least $E_{\text{cutoff}} \approx 200$ MeV. Such an extension improves completeness of the 2p-2h basis and resulting collectivity and the spectral distribution of low‑lying modes, giving better agreement with experiment and stabilizing the energy moments. 
Increasing the cut-off beyond this point, however, produces unphysical high-lying states in RSSTDA, some of which exhibit anomalously large transition strengths, with magnitude ranging from $10^{-3}$ to $10^{-2}$ e\( ^2 \)MeVfm\( ^4 \).
They typically appear above the excitation energy associated with the chosen 2p-2h cutoff value, and are well separated from the rest of excitation spectrum. 
As these states do not correspond to physical excitations, they may be interpreted as spurious artefacts 
and must be excluded from the analysis.
Although the anomalous states contribute no more than a few percent to the $m_0$ moment in the ISGQR for 2p-2h  $E_{cutoff}\gtrsim180$ MeV, their impact on higher-order energy moments is substantial. In particular, the $m_2$ and $m_3$ moments exhibit a pronounced increase, often by several times, when these high-lying states are included in the analysis.
An analogous behavior was observed in the case of the GMR, but the contribution of these states to the \( m_0 \) moment remained below 0.5\% even for relatively large cutoff $150$ MeV $\lesssim E_{cutoff}\lesssim210$ MeV, and they were therefore not excluded from the moment calculations.
Similar effects have been observed in other models including 2p–2h configurations, such as PVC \cite{Litvinova2011}, where high-lying fragmented states, though weak in strength, cause an unphysical enhancement of higher-order energy moments due to their location in the upper part of the spectrum. 

Table~\ref{tabISQGRN14} summarizes the energy moments and the energy‑weighted sum rule (EWSR) for the isoscalar $2^{+}$ transition in $^{16}$O, calculated with the RTDA, RSSTDA(d) and RSSTDA using the DD‑PC1 interaction. Only the positive part of the excitation spectrum ($E_{x} \le 190$ MeV) is considered, and a 2p–2h energy cut‑off $E_{\text{cutoff}} = 210$ MeV is applied for the SSTDA calculations. The zeroth‑moment $m_{0}$, which measures the total transition strength, varies only modestly among the three models, i.e., $m_0$ values are 337.849, 325.334 and 324.976 e$^{2}$fm$^{4}$ for RTDA, RSSTDA(d) and RSSTDA, respectively. Within the experimental window, the strength drops markedly for the subtraction‑based models (289.786 e$^{2}$fm$^{4}$ in RSSTDA(d) and 293.582e$^{2}$fm$^{4}$ in RSSTDA), while it remains essentially unchanged in TDA (336.561 e$^{2}$fm$^{4}$), indicating a stronger sensitivity of RSSTDA and RSSTDA(d) to low-energy part of excitation spectrum. The first energy-weighted moment $m_{1}$ increases significantly with the inclusion of 2p–2h configurations. The RTDA calculation gives $m_{1} = 8.07 \times 10^{3}$\,e$^{2}$MeVfm$^{4}$, while the corresponding values obtained in RSSTDA(d) and RSSTDA are $8.94 \times 10^{3}$\,e$^{2}$MeVfm$^{4}$ and $9.19 \times 10^{3}$\,e$^{2}$MeVfm$^{4}$, respectively. This enhancement arises through the subtraction procedure, which eliminates the double counting of static correlations. The resulting upward shift of the transition strength emphasizes the essential role of a consistent subtraction scheme in preserving the physical reliability of the model and ensuring an accurate description of the nuclear excitation spectrum.

The centroid energy ($m_1/m_0$) confirms that all theoretical models overestimate the experimental value of $19.76 \pm 0.22$\,MeV\,\cite{Lui2001}. RTDA yields a centroid of 23.899\,MeV, while RSSTDA(d) and RSSTDA predict even higher values of 27.465\,MeV and 28.287\,MeV, respectively. When restricted to the experimental energy window, the centroids are shifted downward, resulting in 23.709 MeV for RTDA, 21.604\,MeV for RSSTDA(d), and 23.008\,MeV for RSSTDA, bringing the subtraction-based models into closer agreement with experiment. The effective transition energy $\sqrt{m_{1}/m_{-1}}$ follows the same trend, with values of 23.771\,MeV in RTDA, 25.055\,MeV in RSSTDA(d), and 25.413\,MeV in RSSTDA. When restricted to the experimental energy window, these values decrease to 23.649\,MeV, 21.431\,MeV, and 22.290\,MeV, respectively, confirming that the subtraction procedure shifts a significant portion of the strength toward lower excitation energies. It is also worth noting that the inverse energy-weighted moment $m_{-1}$ remains essentially conserved between the RSSTDA and RSSTDA(d) calculations when the entire positive part of the excitation spectrum is included, including the high-lying anomalous states above 180\,MeV. However, when the analysis is restricted to the spectrum below 190\,MeV, a small difference between the two models does appear, indicating a residual sensitivity of $m_{-1}$ to the redistribution of low-energy strength. 
Larger increase appears in the ratio $\sqrt{m_{3}/m_{1}}$, which reaches 63.975 MeV in RSSTDA(d) and 59.911 MeV in RSSTDA, compared with only 26.432 MeV in RTDA, highlighting the role of 2p–2h configurations in populating the high‑energy tail of the spectrum. When restricted to the experimental energy window, these values are significantly reduced, to 22.502\,MeV in RSSTDA(d) and 25.782\,MeV in RSSTDA.

The width of the strength distribution $\Delta$ remains within the same order of magnitude for all models when compared to the experimental value of $5.11 \pm 0.17$\,MeV. Specifically, it amounts to 4.279\,MeV in RTDA, it reduces to 3.347\,MeV in RSSTDA(d), and increases to 6.381\,MeV in RSSTDA. 
Despite these variations, all three approaches provide a qualitatively consistent description of the experimental width. A similar trend is observed in the exhaustion of the energy-weighted sum rule (EWSR). TDA saturates 89.22\,\% of the classical sum rule \cite{Hartmann2002}, RSSTDA(d) reaches 98.73\,\%, and RSSTDA slightly overshoots full saturation with 101.58\,\%. When the analysis is restricted to the experimental energy window, the exhaustion remains nearly unchanged in RTDA (88.17\,\%), but decreases significantly in the subtraction-based models, yielding 69.18\,\% for RSSTDA(d) and 74.64\,\% for RSSTDA. These values are in reasonable agreement with the experimentally observed quenching of $(53 \pm 10)$\,\% \cite{Lui2001}, indicating that the inclusion of 2p–2h correlations, combined with a proper subtraction procedure, leads to a more realistic redistribution of strength across the excitation spectrum.

Although not explicitly shown in the Tab.~\ref{tabISQGRN14}, it is worth noting that for the ISGQR, the RSTDA calculation with a 2p–2h cutoff of 210 MeV preserves the \( m_0 \) moment, while the \( m_1 \) moment is found to be only $\approx0.6\%$ larger than in the corresponding RTDA result. This indicates that the energy-weighted sum rule is nearly fully preserved between the RSTDA and RTDA approaches. The same holds for the RSTDA(d) calculation, whose values differ from those of RSTDA by less than 0.1\%.

\begin{table}
\begin{center}
    \caption{Energy moments and energy weighted sum rule (EWSR) of isoscalar $2^+$  transitions obtained by the RTDA, RSTDA and RSSTDA calculations for $^{16}$O with DD-PC1 interaction. For RSTDA and RSSTDA calculations the energy 2p-2h cutoff is taken $E_{cutoff}=210$ MeV. The width $\Delta$ and values in round brackets are calculated within experimental window from \cite{Lui2001}.}
    \begin{tabular}{c c c c c}
        \hline\hline
         $ $ & RTDA & RSSTDA(d) & RSSTDA & Exp. \cite{Lui2001}\\
        \hline
        \multicolumn{5}{c}{$2^{+}$} \\
        $m_0$ [e$^2$fm$^4$] & 337.849 (336.561) &  325.334 (289.786) & 324.976 (293.582) &   \\
        $m_1$ [e$^2$MeVfm$^4$]  & 8074.100 (7979.671) & 8935.411 (6260.406) & 9192.603 (6754.598) &  \\
        $m_{-1}$ [e$^2$MeV$^{-1}$fm$^4$] & 14.288 (14.268) & 14.234 (13.631) & 14.234 (13.595) & \\
        $m_1/m_0$[MeV] & 23.899 (23.709) & 27.465 (21.604) & 28.287 (23.008) & 19.76$\pm$0.22\\      
        $\sqrt{m_1/m_{-1}}$ [MeV] & 23.771 (23.649)& 25.055 (21.431) & 25.413 (22.290)  & \\
        $\sqrt{m_3/m_{1}}$ [MeV] & 26.432 (23.961)& 63.975 (22.502) & 59.911 (25.782)  & \\
        $\Delta$ [MeV] & 4.279 & 3.347 & 6.381 & 5.11$\pm$ 0.17 \\
        EWSR[\%] & 89.22 (88.17) & 98.73 (69.18) & 101.58 (74.64) & 53$\pm$10 \\
        \hline
    \end{tabular}
    \label{tabISQGRN14}
\end{center}
\end{table}

\begin{figure}
  \includegraphics[width=0.9\linewidth]{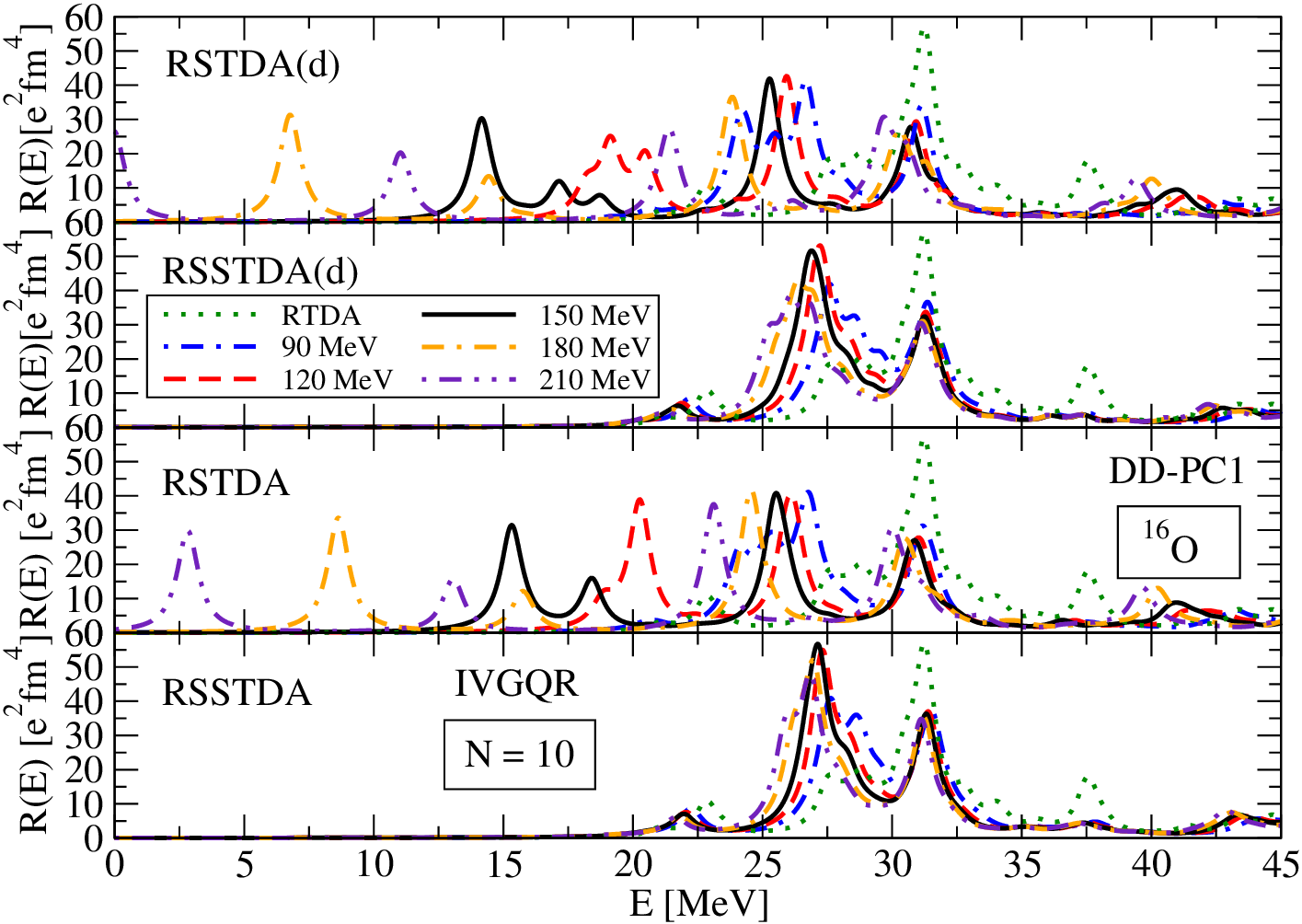}
  \caption{The same as Fig. \ref{fig:ISGMR:N20}, but for IVGQR strength distribution.}
  \label{fig:IVGQR:N14}
\end{figure}

\begin{figure}
  \includegraphics[width=0.9\linewidth]{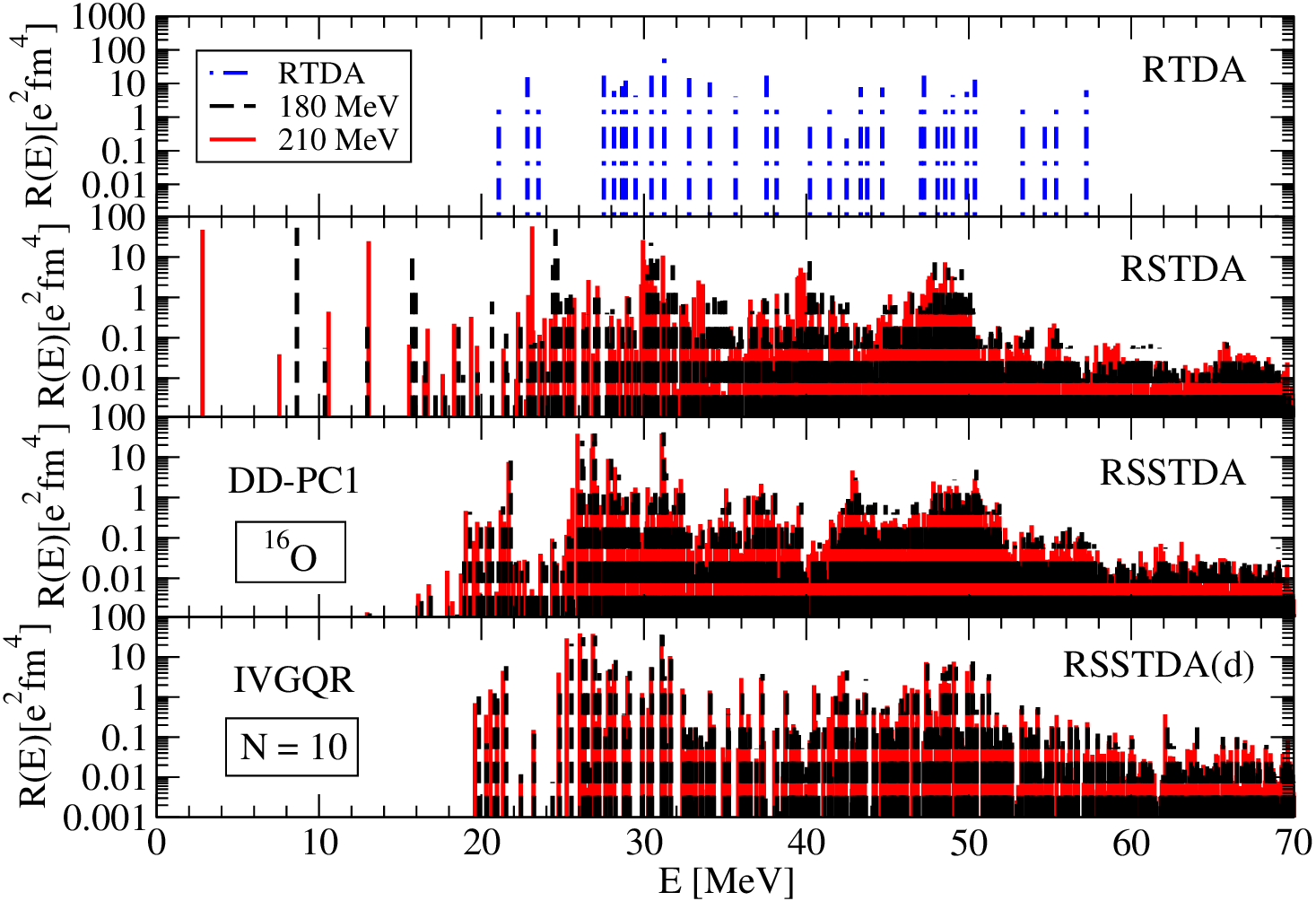}
  \caption{The same as Fig. \ref{fig:ISGMR:N20:Discrete}, but for IVGQR strength distribution.}
  \label{fig:IVGQR:N14:discrete}
\end{figure}

Finally, we also study the isovector quadrupole response within the RSTDA established in this work. Figures~\ref{fig:IVGQR:N14} and \ref{fig:IVGQR:N14:discrete} show the Lorentz-smoothed and discrete IVGQR strength distributions of $^{16}$O, respectively. The results for the IVGQR are qualitatively similar to those for the ISGQR, except that the difference between the RSTDA and RSTDA(d) spectra is considerably smaller in the IVGQR case. The same trend is observed in the subtracted version, possibly indicating a less pronounced role of the interaction among 2p–2h configurations in shaping the IVGQR strength. This is further supported by the weak energy dependence on the 2p–2h cutoff energy observed in RSSTDA and RSSTDA(d), where the results tend to converge for $E_x \gtrsim 150$ MeV.

\subsection{Cumulative strength of GMR and GQR}

\begin{figure}
  \includegraphics[width=0.9\linewidth]{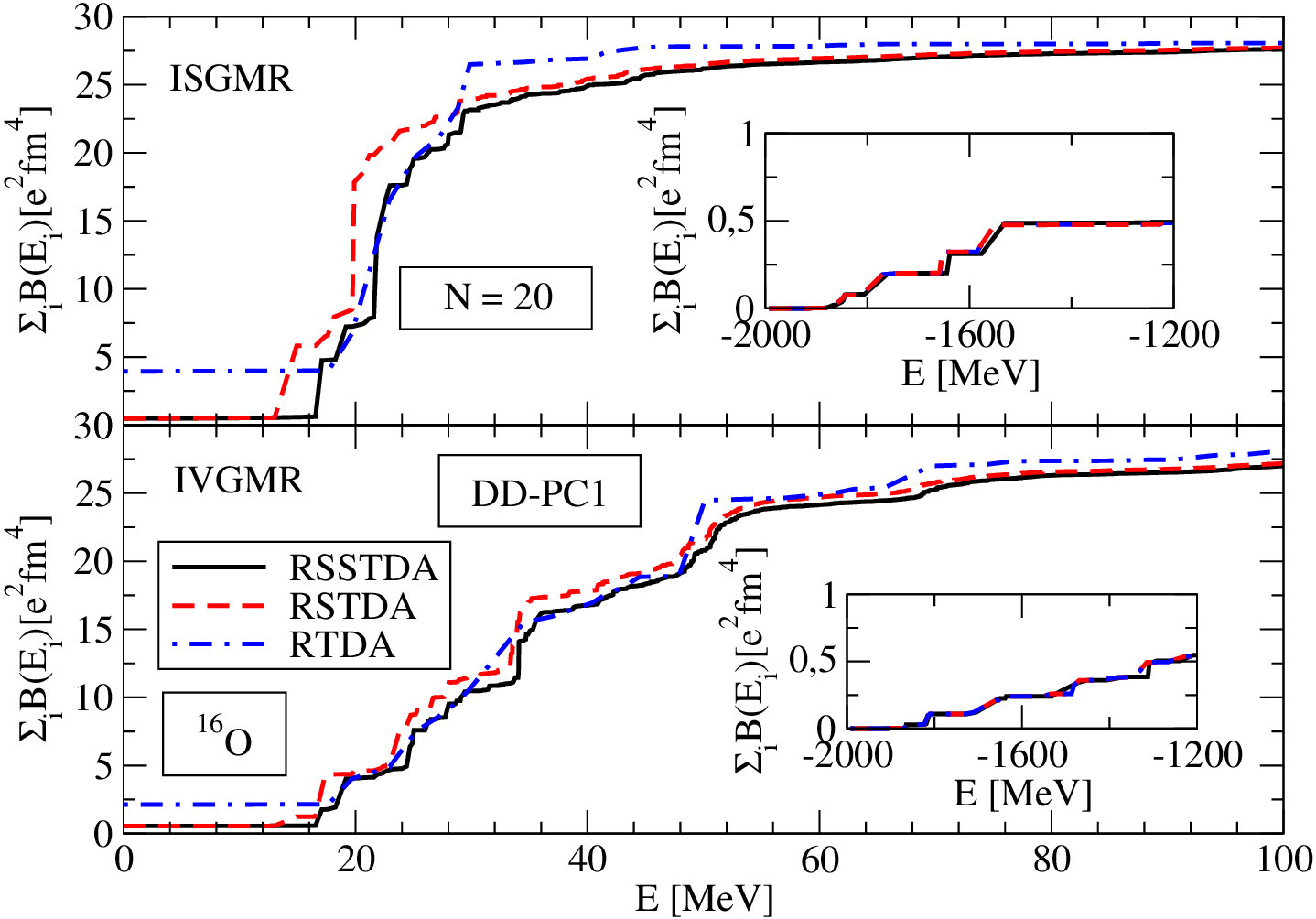}
  \caption{Cummulative strength distributions for ISGMR and IVGMR in $^{16}$O for RTDA, RSTDA and RSSTDA responses as functions of excitation energy 2p-2h cutoff $E_{cutoff} = 150$ MeV.}
  \label{fig:GMR:N20:CUM}
\end{figure}

In the following we consider the cumulative transition 
strength $\sum_i B(E_i)$ for the monopole and quadrupole transitions.
Figure \ref{fig:GMR:N20:CUM} presents the calculated values of $\sum_i B(E_i)$ for ISGMR and IVGMR in $^{16}$O as a function of excitation energy $E$, using 2p-2h energy cutoff $E_{cutoff} = 150$ MeV. Three theoretical approaches, i.e., RTDA, RSTDA, and RSSTDA calculations with the DD-PC1 interaction are compared. The energy spectrum extends up to 100 MeV (the energy part $E_x > 100$ MeV is excluded in Fig. \ref{fig:GMR:N20:CUM}), with RSSTDA yielding a notably larger amount of strength at higher energies compared to RSTDA and RTDA.  
At negative part of spectrum ($[-2000,~-1200]$ MeV), all three methods produce similar results, but deviations become more pronounced at positive part of the energy spectrum, typically after $E_x\gtrsim10$ MeV, emphasizing the role of 2p-2h correlations in RSTDA and RSSTDA.  
The cumulative ISGMR strength saturates more rapidly in RTDA and RSTDA in the energy region $ 10\lesssim E_x \lesssim40$ MeV, while RSSTDA exhibits a broader distribution. For $E_x \gtrsim 40$~MeV, RSTDA and RSSTDA start to converge to similar cumulative strength values, both of which remain slightly lower than RTDA.
This is in correlation with the moment ratio \( \sqrt{m_1/m_{-1}} \) in RSSTDA, which is higher compared to the experimental value, indicating a redistribution of strength toward higher excitation energies (see Tab. \ref{tabIS}). This behavior is further supported by the ratio \( \sqrt{m_3/m_1} \), which is somewhat larger in both RSTDA and RSSTDA compared to the experimental value, suggesting an enhanced contribution of high-energy excitations. Notably, the ISGMR and IVGMR spectra reveal different response characteristics, with ISGMR exhibiting a smoother increase of cumulative sum and IVGMR showing a more fragmented strength distribution. In general, the cumulative transition strength in RSTDA for the IVGMR increases more rapidly than in RSSTDA, however, they tend to converge to similar values above $E_x \gtrsim 65$~MeV.  

\begin{figure}
  \includegraphics[width=0.9\linewidth]{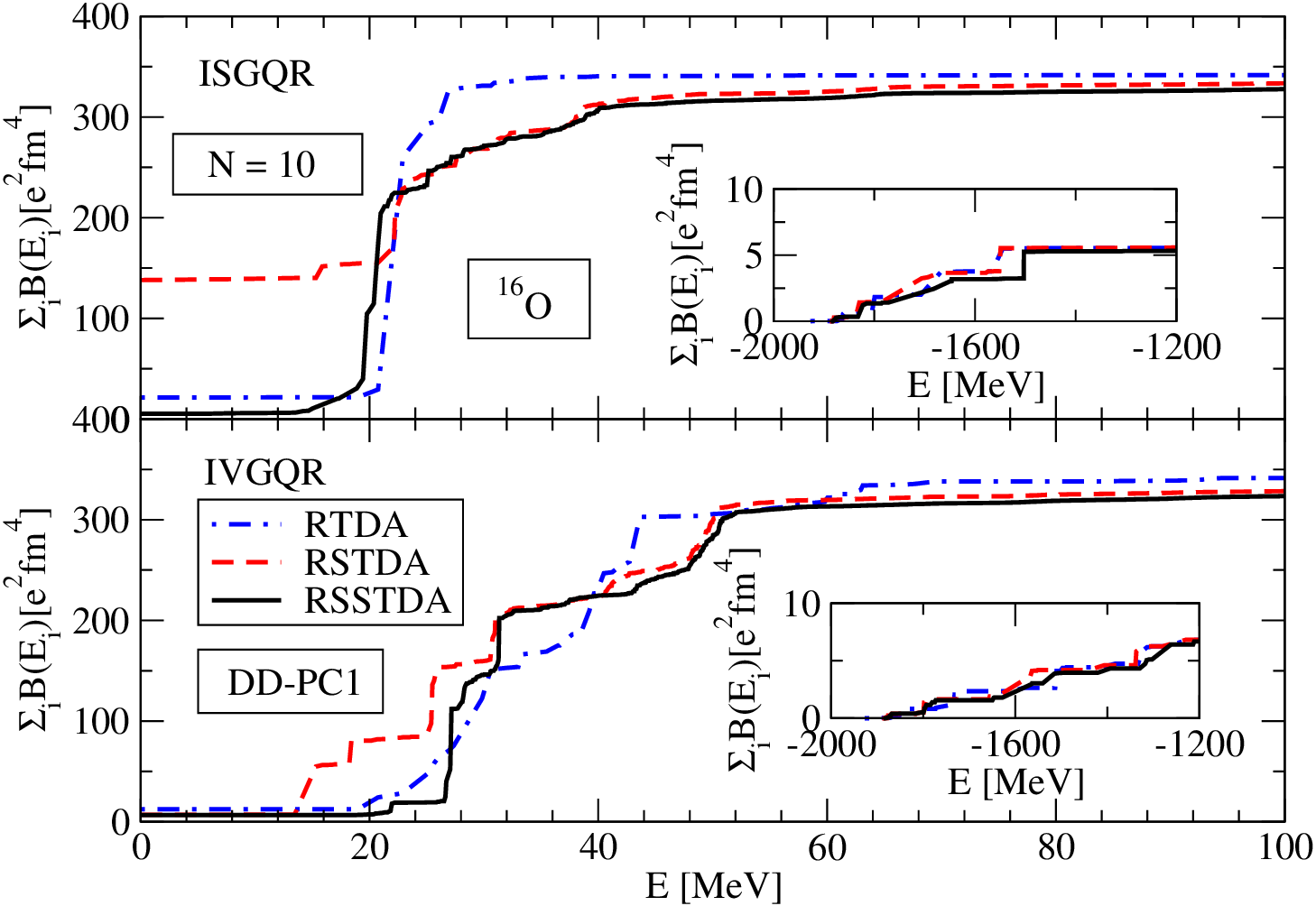}
  \caption{The same as Fig. \ref{fig:GMR:N20:CUM}, but for ISGQR and IVGQR.}
  \label{fig:GQR:N14:CUM}
\end{figure}

The cumulative transition strength $\sum_i B(E_i)$ for the ISGQR and IVGQR in $^{16}$O, obtained using RTDA, RSTDA, and RSSTDA, are shown in Fig.~\ref{fig:GQR:N14:CUM}. In the ISGQR case, the RTDA strength rises sharply in the region $20 \lesssim E_x \lesssim 30$~MeV, reaching more than 95\% of the total strength. In other words, RTDA exhibits the most rapid saturation, followed by RSTDA, while RSSTDA accumulates strength more gradually over a broader energy range. This behavior reflects the impact of 2p--2h correlations and the subtraction procedure in RSSTDA, which redistributes the strength toward higher excitation energies. The convergence of the RSTDA and RSSTDA curves at higher energies suggests that their differences are most pronounced in the low-to-intermediate energy region. This difference becomes even more pronounced in the IVGQR case, where the cumulative strength in RSSTDA significantly deviates from RSTDA in the region $E_x \lesssim 45$~MeV, emphasizing the enhanced role of correlations in the isovector response. The RSTDA cumulative strength increases more rapidly in the lower part of the spectrum compared to RSSTDA and RTDA. This suggests that the subtraction method employed in RSSTDA effectively removes double counting of correlations and restores spectral stability, thereby allowing a more physically consistent redistribution of transition strength toward higher energies, similar to the ISGQR case.

\subsection{Microscopic distinction of RTDA, RSTDA and RSSTDA}\label{Micro}

In the RTDA calculation of ISGMR only 1p-1h configurations contribute to the transition strength. As an example of an eigenstate bellow 20 MeV, the contributions of 1p-1h pairs at 17.67 MeV is shown in Tab. \ref{TDA:TabEn}.  In this state, the transition strength is highly concentrated in a single dominant configuration, $(2p_{1/2})^\pi (1p_{1/2})^\pi$, which contributes 98.54\% of the total norm. The remaining particle-hole (p-h) pairs contribute significantly less, with only one additional configuration exceeding 1\% and most others contributing below 0.001.
In the case of RSTDA calculations of ISGMR, the states below 21 MeV that contribute with $B \gtrsim 1.0 \, e^2 \mathrm{fm}^4$ are characterized by strong mixing of 1p-1h and 2p-2h configurations, occurring within the range of approximately 20\% to 80\% in favor of 2p-2h configurations. This does not include few states that are generally characterized by $B \lesssim 1.0 \, e^2 \mathrm{fm}^4$, predominantly of 2p-2h nature, which can exhibit 2p-2h configuration contributions exceeding 99\%. 

An example of the RSTDA state, which may be considered as a counterpart of the previously discussed RTDA state at $17.67$ MeV, is the eigenvalue at $14.93$ MeV, with its structure presented in Tab.~\ref{STDA14p93}. The $(2p_{1/2})^\pi (1p_{1/2})^\pi$ configuration clearly dominates the p–h part with a contribution of $0.68272$, followed by $(2p_{1/2})^\nu (1p_{1/2})^\nu$ with $0.10516$, while all remaining p–h configurations contribute less than $0.04$. In the 2p–2h part, the largest individual amplitude is $0.01043$, and only one configuration exceeds $0.01$, whereas 27 configurations lie in the range $[0.001$–$0.01]$, and over $7300$ configurations have contributions smaller than $0.001$. Although no dominant 2p–2h configuration emerges, the cumulative contribution from many weakly coupled components adds up to a significant value. This indicates that correlations beyond the simple p–h picture play a non-negligible role even in states that appear to be predominantly of p–h nature. The presence of such a broad distribution of small 2p–2h components contributes to the fragmentation and spreading of the strength in RSTDA to lower excitation energies.

It now remains to clarify the role of the subtraction method in shaping the microscopic structure of the eigenstates within the investigated energy range. Compared to RSTDA, RSSTDA induces a systematic upward shift of excitation energies and alters the internal configuration mixing. 
This behavior is consistent with the theoretical foundation of the subtraction method, which eliminates the static components of ground-state correlations. As a result, RSSTDA suppresses the artificial enhancement of 2p–2h admixtures in low-energy mixed configurations and favors more physical 1p–1h–dominated structures. In states where 2p–2h components are still prominent, the subtraction does not remove them entirely but rather reshapes their distribution, concentrating the strength into fewer, more collective configurations and reducing fragmentation. The structure of predominantly 2p–2h states with $N_2\gtrsim0.99$ remains largely unaffected. Consequently, the method stabilizes the low-energy spectrum and mitigates the spreading of transition strength toward lower excitation energies, as observed in the comparison of counterpart states in RSTDA and RSSTDA.

The state that represents the counterpart of the one identified in the RSTDA calculation at $14.93$ MeV (shown in Tab.~\ref{STDA14p93}), which appears at $17.09$ MeV within the RSSTDA approximation, is presented in Tab.~\ref{SSTDA17p09}. The dominant particle-hole (p–h) configuration, $(2p_{1/2})^\pi (1p_{1/2})^\pi$, significantly increases in strength from $0.68272$ in RSTDA to $0.84964$ in RSSTDA, indicating a notable reduction in configuration mixing. Simultaneously, the contribution of 2p–2h configurations is considerably suppressed. The leading 2p–2h amplitude in RSSTDA reaches only $0.02149$, while in RSTDA it amounted to $0.01043$, with additional configurations exceeding $0.01$. The total 2p–2h norm in RSSTDA is approximately $N_2 \approx 0.05$, suggesting a redistribution and partial suppression of correlation effects beyond the p–h space. This transformation is reflected in the systematic upward shift of excitation energies (below $21$ MeV) by several MeV relative to the RSTDA calculation with 2p–2h cutoff $E_{\text{cutoff}} = 150$ MeV. When taking into account a few more states of similar character in the energy spectrum, this behavior leads to the overall suppression of low-energy fragmentation effects in RSSTDA.

\begin{table}[h]
\begin{center}
    \caption{Particle-hole (p-h) configurations with their contribution in the RTDA case for the eigenvalue at $E_x = 17.67$ MeV.}
    \begin{tabular}{cc}
        \hline\hline
        Configuration & Value \\
        \hline
        $(2p_{1/2})^\pi$ $(1p_{1/2})^\pi$ & 0.98540 \\
        $(2p_{1/2})^\nu$ $(1p_{1/2})^\nu$ & 0.01151 \\
        $(2p_{3/2})^\nu$ $(1p_{3/2})^\nu$ & 0.00126 \\
        $(3p_{1/2})^\nu$ $(1p_{1/2})^\nu$ & 0.00085 \\
        $(3p_{3/2})^\nu$ $(1p_{3/2})^\nu$ & 0.00048 \\
        \hline
        Number of ph configurations: & 108 \\
        Range [0.1 - 0.9999] & 1 \\
        Range [0.01 - 0.1] & 1 \\
        Range [0.001 - 0.01] & 1 \\
        Range [< 0.001] & 105 \\
        \hline\hline
    \end{tabular}
    \label{TDA:TabEn}
\end{center}
\end{table}

\begin{table}[h]
\begin{center}
    \caption{Particle-hole (p–h) and two-particle–two-hole (2p–2h) configurations with their contribution in the RSTDA calculation for the eigenvalue at $E_x = 14.93$ MeV.}
    \begin{tabular}{cc}
        \hline\hline
        Configuration & Value \\
        \hline
        $(2p_{1/2})^\pi$ $(1p_{1/2})^\pi$ & 0.68272 \\
        $(2p_{1/2})^\nu$ $(1p_{1/2})^\nu$ & 0.10516 \\
        $(3p_{1/2})^\pi$ $(1p_{1/2})^\pi$ & 0.03856 \\
        $(3p_{1/2})^\nu$ $(1p_{1/2})^\nu$ & 0.01524 \\
        $(4p_{1/2})^\pi$ $(1p_{1/2})^\pi$ & 0.00789 \\
        $(4p_{1/2})^\nu$ $(1p_{1/2})^\nu$ & 0.00336 \\
        $(5p_{1/2})^\pi$ $(1p_{1/2})^\pi$ & 0.00208 \\
        $(2s_{1/2})^\pi$ $(1s_{1/2})^\pi$ & 0.00198 \\
        \hline
        Number of p–h configurations: & 108 \\
        Range [0.1 - 0.9999] & 2 (0.78787) \\
        Range [0.01 - 0.1] & 2 (0.05380) \\
        Range [0.001 - 0.01] & 4 (0.01531) \\
        Range [< 0.001] & 100 (0.00791) \\
        \hline
        $[(2s_{1/2},2s_{1/2})^{0}_{\pi}(1p_{1/2},1p_{1/2})^{0}_{\pi}]^{0}$ & 0.01043 \\
        $[(2s_{1/2},2s_{1/2})^{0}_{\nu}(1p_{1/2},1p_{1/2})^{0}_{\nu}]^{0}$ & 0.00537 \\
        $[(2p_{1/2},1d_{5/2})^{2}_{\pi}(1s_{1/2},1p_{3/2})^{2}_{\pi}]^{0}$ & 0.00438 \\
        $[(2s_{1/2},1d_{3/2})^{2}_{\pi}(1p_{3/2},1p_{3/2})^{2}_{\pi}]^{0}$ & 0.00333 \\
        $[(2s_{1/2},1d_{5/2})^{2}_{\pi}(1p_{3/2},1p_{3/2})^{2}_{\pi}]^{0}$ & 0.00325 \\
        $[(1d_{3/2},1d_{5/2})^{2}_{\pi}(1p_{3/2},1p_{3/2})^{2}_{\pi}]^{0}$ & 0.00318 \\
        $[(3p_{1/2},1d_{5/2})^{2}_{\pi}(1s_{1/2},1p_{3/2})^{2}_{\pi}]^{0}$ & 0.00299 \\
        $[(2s_{1/2},2p_{1/2})^{0}_{\pi}(1s_{1/2},1p_{1/2})^{0}_{\pi}]^{0}$ & 0.00294 \\
        $[(2s_{1/2},1d_{5/2})^{2}_{\nu}(1p_{3/2},1p_{3/2})^{2}_{\nu}]^{0}$ & 0.00289 \\
        $[(1d_{3/2},1d_{5/2})^{2}_{\nu}(1p_{3/2},1p_{3/2})^{2}_{\nu}]^{0}$ & 0.00260 \\
        \hline
        Number of 2p–2h configurations: & 7421 \\
        Range [0.01 - 0.1] & 1 (0.01043) \\
        Range [0.001 - 0.01] & 27 (0.06071) \\
        Range [< 0.001] & 7393 (0.06397) \\
        \hline\hline
    \end{tabular}
    \label{STDA14p93}
\end{center}
\end{table}

\begin{table}[h]
\begin{center}
    \caption{Particle-hole (p–h) and two-particle–two-hole (2p–2h) configurations with their contribution in the RSSTDA calculation for the eigenvalue at $E_x = 17.09$ MeV.}
    \begin{tabular}{cc}
        \hline\hline
        Configuration & Value \\
        \hline
        $(2p_{1/2})^\pi$ $(1p_{1/2})^\pi$ & 0.84964 \\
        $(2p_{1/2})^\nu$ $(1p_{1/2})^\nu$ & 0.03654 \\
        $(3p_{1/2})^\pi$ $(1p_{1/2})^\pi$ & 0.00349 \\
        $(3p_{1/2})^\nu$ $(1p_{1/2})^\nu$ & 0.00278 \\
        $(2p_{3/2})^\nu$ $(1p_{3/2})^\nu$ & 0.00160 \\
        \hline
        Number of p–h configurations: & 108 \\
        Range [0.1 - 0.9999] & 1 (0.84964) \\
        Range [0.01 - 0.1] & 1 (0.03654) \\
        Range [0.001 - 0.01] & 3 (0.00788) \\
        Range [< 0.001] & 103 (0.00294) \\
        \hline
        $[(2s_{1/2},2s_{1/2})^{0}_{\pi}(1p_{1/2},1p_{1/2})^{0}_{\pi}]^{0}$ & 0.02149 \\
        $[(2s_{1/2},2s_{1/2})^{0}_{\nu}(1p_{1/2},1p_{1/2})^{0}_{\nu}]^{0}$ & 0.00996 \\
        $[(2p_{1/2},1d_{5/2})^{2}_{\pi}(1s_{1/2},1p_{3/2})^{2}_{\pi}]^{0}$ & 0.00339 \\
        $[(2s_{1/2},2p_{1/2})^{0}_{\pi}(1s_{1/2},1p_{1/2})^{0}_{\pi}]^{0}$ & 0.00240 \\
        $[(3p_{1/2},1d_{5/2})^{2}_{\pi}(1s_{1/2},1p_{3/2})^{2}_{\pi}]^{0}$ & 0.00227 \\
        $[(2s_{1/2},3s_{1/2})^{0}_{\pi}(1p_{1/2},1p_{1/2})^{0}_{\pi}]^{0}$ & 0.00211 \\
        $[(2p_{1/2},3p_{3/2})^{1}_{\pi}(1p_{1/2},1p_{3/2})^{1}_{\pi}]^{0}$ & 0.00147 \\
        $[(1d_{5/2},2d_{5/2})^{2}_{\pi}(1p_{1/2},1p_{3/2})^{2}_{\pi}]^{0}$ & 0.00144 \\
        $[(1d_{5/2},2d_{5/2})^{2}_{\pi}(1p_{3/2},1p_{3/2})^{2}_{\pi}]^{0}$ & 0.00141 \\
        $[(2p_{1/2},4p_{3/2})^{1}_{\pi}(1p_{1/2},1p_{3/2})^{1}_{\pi}]^{0}$ & 0.00139 \\
        \hline
        Number of 2p–2h configurations: & 7421 \\
        Range [0.01 - 0.1] & 1 (0.02149) \\
        Range [0.001 - 0.01] & 13 (0.03069) \\
        Range [< 0.001] & 7407 (0.05082) \\
        \hline\hline
    \end{tabular}
    \label{SSTDA17p09}
\end{center}
\end{table}

Such changes also occur in the intermediate excitation energy range, approximately between 20 and 35~MeV, where the differences between RSTDA and RSSTDA typically lead to variations in the mixing between 1p--1h and 2p--2h configurations ranging from a few percent up to about 20\%, depending on the specific state. As an example of such counterpart states that contribute most to the transition strength, see Tabs.~\ref{STDA19p88} and \ref{SSTDA21p84} for the RSTDA and RSSTDA calculations, respectively. 
In the RSTDA case, the dominant state is located at $19.88$ MeV. The $(2p_{3/2})^\pi (1p_{3/2})^\pi$ and $(2p_{3/2})^\nu (1p_{3/2})^\nu$ configurations together contribute $0.37614$ to the norm, while seven additional 1p–1h configurations fall in the $[0.01$–$0.1]$ range with a total of $0.19378$. Thus, the total contribution of the 1p-1h configuration accounts for approximately $59\%$ of the state norm. The dominant 2p–2h configuration in RSTDA reaches $0.02687$, with six more contributing above $0.01$, and a large number of small components below $0.01$ summing to an additional $0.18403$.

In the RSSTDA calculation, the counterpart state is found at $21.84$ MeV. The same $(2p_{3/2})^\pi (1p_{3/2})^\pi$ and $(2p_{3/2})^\nu (1p_{3/2})^\nu$ configurations now contribute $0.28416$ and $0.09317$, respectively, giving a total of $0.37733$. Although this is similar to the RSTDA value, the $N_1$ norm remains slightly lower, with fewer configurations above $0.01$ and more in the subdominant range. On the other hand, the 2p--2h part becomes more concentrated in RSSTDA, as the leading configuration $[(1d_{3/2},1d_{5/2})^{1}_{\pi} (1p_{1/2},1p_{3/2})^{1}_{\pi}]^{0}$ contributes $0.10138$, followed by $[(1d_{3/2},1d_{5/2})^{1}_{\nu}(1p_{1/2},1p_{3/2})^{1}_{\nu}]^{0}$ with $0.08002$, while the remaining seven configurations above $0.01$ together contribute only $0.09307$. This redistribution leads to a slightly larger overall contribution from the most collective 2p--2h components in RSSTDA compared to RSTDA, with the total norm of the 2p--2h sector increasing from approximately $0.41$ in RSTDA to about $0.53$ in RSSTDA.

These results indicate that at intermediate energies, the subtraction method implemented in RSSTDA does not drastically change the total 2p–2h content but rather reshapes its distribution. The strength becomes concentrated in fewer but more collective 2p–2h configurations, while the fragmentation is reduced compared to RSTDA. This pattern supports the interpretation that the subtraction method primarily affects the structure and mixing of correlations rather than eliminating them entirely.

\begin{table}[h]
\begin{center}
    \caption{Particle-hole (p–h) and two-particle–two-hole (2p–2h) configurations with their contribution in the RSTDA calculation for the eigenvalue at $E_x = 19.88$ MeV.}
    \begin{tabular}{cc}
        \hline\hline
        Configuration & Value \\
        \hline
        $(2p_{3/2})^\pi$ $(1p_{3/2})^\pi$ & 0.25752 \\
        $(2p_{3/2})^\nu$ $(1p_{3/2})^\nu$ & 0.11862 \\
        $(3p_{3/2})^\pi$ $(1p_{3/2})^\pi$ & 0.04796 \\
        $(3p_{3/2})^\nu$ $(1p_{3/2})^\nu$ & 0.03889 \\
        $(2s_{1/2})^\nu$ $(1s_{1/2})^\nu$ & 0.03679 \\
        $(2s_{1/2})^\pi$ $(1s_{1/2})^\pi$ & 0.03547 \\
        $(4p_{3/2})^\pi$ $(1p_{3/2})^\pi$ & 0.01226 \\
        $(3s_{1/2})^\pi$ $(1s_{1/2})^\pi$ & 0.01197 \\
        $(4p_{3/2})^\nu$ $(1p_{3/2})^\nu$ & 0.01043 \\
        \hline
        Number of p–h configurations: & 108 \\
        Range [0.1 - 0.9999] & 2 (0.37614) \\
        Range [0.01 - 0.1] & 7 (0.19378) \\
        Range [0.001 - 0.01] & 7 (0.01849) \\
        Range [< 0.001] & 92 (0.00612) \\
        \hline
        $[(2s_{1/2},2s_{1/2})^{0}_{\pi}(1p_{1/2},1p_{1/2})^{0}_{\pi}]^{0}$ & 0.02687 \\
        $[(3s_{1/2},1d_{3/2})^{2}_{\pi}(1p_{1/2},1p_{3/2})^{2}_{\pi}]^{0}$ & 0.02521 \\
        $[(3s_{1/2},1d_{3/2})^{2}_{\nu}(1p_{1/2},1p_{3/2})^{2}_{\nu}]^{0}$ & 0.01952 \\
        $[(3s_{1/2},1d_{3/2})^{2}_{\pi}(1p_{3/2},1p_{3/2})^{2}_{\pi}]^{0}$ & 0.01670 \\
        $[(3s_{1/2},1d_{3/2})^{2}_{\nu}(1p_{3/2},1p_{3/2})^{2}_{\nu}]^{0}$ & 0.01340 \\
        $[(2s_{1/2},1d_{3/2})^{1}_{\pi}(1p_{1/2},1p_{3/2})^{1}_{\pi}]^{0}$ & 0.01232 \\
        $[(2s_{1/2},2s_{1/2})^{0}_{\nu}(1p_{1/2},1p_{1/2})^{0}_{\nu}]^{0}$ & 0.01180 \\
        $[(2s_{1/2},1d_{3/2})^{1}_{\nu}(1p_{1/2},1p_{3/2})^{1}_{\nu}]^{0}$ & 0.00947 \\
        $[(2s_{1/2},1d_{5/2})^{2}_{\pi}(1p_{3/2},1p_{3/2})^{2}_{\pi}]^{0}$ & 0.00739 \\
        $[(2s_{1/2},3s_{1/2})^{1}_{\nu}(1p_{1/2},1p_{3/2})^{1}_{\nu}]^{0}$ & 0.00688 \\
        \hline
        Number of 2p–2h configurations: & 7421 \\
        Range [0.01 - 0.1] & 7 (0.12582) \\
        Range [0.001 - 0.01] & 54 (0.18403) \\
        Range [< 0.001] & 7360 (0.09564) \\
        \hline\hline
    \end{tabular}
    \label{STDA19p88}
\end{center}
\end{table}

\begin{table}[h]
\begin{center}
    \caption{Particle-hole (p–h) and two-particle–two-hole (2p–2h) configurations with their contribution in the RSSTDA calculation for the eigenvalue at $E_x = 21.84$ MeV.}
    \begin{tabular}{cc}
        \hline\hline
        Configuration & Value \\
        \hline
        $(2p_{3/2})^\pi$ $(1p_{3/2})^\pi$ & 0.28416 \\
        $(2p_{3/2})^\nu$ $(1p_{3/2})^\nu$ & 0.09317 \\
        $(3p_{3/2})^\nu$ $(1p_{3/2})^\nu$ & 0.02073 \\
        $(3p_{3/2})^\pi$ $(1p_{3/2})^\pi$ & 0.02010 \\
        $(2s_{1/2})^\nu$ $(1s_{1/2})^\nu$ & 0.01456 \\
        $(2s_{1/2})^\pi$ $(1s_{1/2})^\pi$ & 0.01341 \\
        \hline
        Number of p–h configurations: & 108 \\
        Range [0.1 - 0.9999] & 1 (0.28416) \\
        Range [0.01 - 0.1] & 5 (0.16197) \\
        Range [0.001 - 0.01] & 6 (0.01705) \\
        Range [< 0.001] & 96 (0.00484) \\
        \hline
        $[(1d_{3/2},1d_{5/2})^{1}_{\pi}(1p_{1/2},1p_{3/2})^{1}_{\pi}]^{0}$ & 0.10138 \\
        $[(1d_{3/2},1d_{5/2})^{1}_{\nu}(1p_{1/2},1p_{3/2})^{1}_{\nu}]^{0}$ & 0.08002 \\
        $[(2s_{1/2},1d_{3/2})^{1}_{\pi}(1p_{1/2},1p_{3/2})^{1}_{\pi}]^{0}$ & 0.01564 \\
        $[(1d_{3/2},1d_{3/2})^{2}_{\pi}(1p_{1/2},1p_{3/2})^{2}_{\pi}]^{0}$ & 0.01413 \\
        $[(1d_{3/2},1d_{3/2})^{2}_{\nu}(1p_{1/2},1p_{3/2})^{2}_{\nu}]^{0}$ & 0.01395 \\
        $[(2s_{1/2},1d_{3/2})^{1}_{\nu}(1p_{1/2},1p_{3/2})^{1}_{\nu}]^{0}$ & 0.01300 \\
        $[(1d_{3/2},1d_{5/2})^{2}_{\nu}(1p_{3/2},1p_{3/2})^{2}_{\nu}]^{0}$ & 0.01298 \\
        $[(1d_{3/2},1d_{5/2})^{2}_{\pi}(1p_{3/2},1p_{3/2})^{2}_{\pi}]^{0}$ & 0.01252 \\
        $[(1d_{3/2},2d_{5/2})^{2}_{\pi}(1p_{1/2},1p_{3/2})^{2}_{\pi}]^{0}$ & 0.01085 \\
        \hline
        Number of 2p–2h configurations: & 7421 \\
        Range [0.1 - 0.9999] & 1 (0.10138) \\
        Range [0.01 - 0.1] & 8 (0.17309) \\
        Range [0.001 - 0.01] & 53 (0.16885) \\
        Range [< 0.001] & 7359 (0.08866) \\
        \hline\hline
    \end{tabular}
    \label{SSTDA21p84}
\end{center}
\end{table}

\section{Conclusions}  

In this work, we introduced the RSTDA based on relativistic nuclar energy density functional with point coupling interaction and performed calculations of the isoscalar and isovector monopole and quadrupole response in $^{16}$O. Our results show qualitative similarities in the energy shifts and fragmentation patterns to those obtained with non-relativistic SRPA employing Skyrme and Gogny interactions, as well as with those using realistic nucleon-nucleon interactions. Specifically, the inclusion of 2p-2h configurations leads to peak fragmentation and shifts to lower excitation energies, which aligns with trends observed in calculations beyond the RTDA level.  
At sufficiently high cutoff values ($E_x\gtrsim120$ MeV), the RSTDA-calculated peaks for the ISGMR move closer to the experimental results, while for the ISGQR, a portion of the strength is shifted into the negative part of the excitation spectrum.
The eigenenergies obtained using the RSSTDA model are systematically higher than those from RSTDA, while their peak positions are up to $\approx 1.5$ MeV lower compared to RTDA. While not all experimentally observed low-lying states are reproduced in the case of ISGMR and ISGQR, the RSSTDA captures the majority of them, including their main structural features, with excitation energies deviating by up to $\approx2$ MeV. Both deviations are partially attributed to the use of the DD-PC1 interaction, which was not originally optimized for describing light nuclei such as $^{16}$O. The second reason for the discrepancy arises from the use of the relativistic Hartree (RH) approximation for the description of the ground state. In the RH approach, no explicit beyond-mean-field correlations are included in the ground-state wavefunction. As a result, certain correlation effects that can influence the excitation spectrum are absent, and this lack of ground-state correlations can significantly affect excitation energies and strength distributions \cite{PhysRevC.87.054330}. Another source of discrepancy, in the case of ISGQR, lies in restriction of the harmonic oscillator basis implemented in the RH model,
using major quantum number $N = 10$ due to slow convergence of RH + RTDA for $^{16}$O with respect to $N$.  Despite these limitations, no additional phenomenological parameters, such as quenching factors, were introduced in both RSTDA and RSSTDA calculations.

Although RSTDA yields results that are in better agreement with experimentally observed values for the ISGMR, its main limitation is complete lack of eigenvalue convergence with increasing 2p–2h cutoff, a deficiency that becomes particularly evident in the case of the ISGQR. On the other hand, it is demonstrated that the RSSTDA calculations converge reasonably well, with results showing stability for both the GMR and GQR. The subtraction method successfully eliminates infrared divergence and double counting while preserving the essential physical features of the response function $R_{RSSTDA}(\omega = 0)\rightarrow R_{RTDA}$. The calculated strength distributions exhibit fragmentation patterns similar to those seen in non-relativistic and realistic interaction-based models, and the behavior of the total strength and EWSR also follows the general trends observed in these frameworks, further supporting the reliability of our approach. Furthermore, for a complete calculation of the positive part of the spectrum, the $m_{-1}$  moment is conserved between RTDA, RSSTDA, and RSSTDA(d), in contrast to previous results obtained with non-relativistic models, where a small discrepancy was observed in SSTDA(d) \cite{Yang2021}. This discrepancy arises from the fact that in non-relativistic models, the diagonal approximation was applied only in the inversion of the \( A_{22} \) matrix, whereas in our relativistic approach it was consistently employed throughout the entire calculation. In contrast to previous non-relativistic implementations, our formulation is fully embedded in the relativistic density functional framework and treats subtraction in a manner that ensures compatibility with the underlying Lagrangian.
Nevertheless, the inclusion of higher-order configurations such as 1p$\text{–}1\alpha\text{–}$2h and $2\alpha\text{–}$2h in the excitation operator remains an open question, as does the consistent subtraction treatment of matrix elements involving antiparticle configurations. A systematic investigation of these effects is planned in a subsequent work.

The observed overestimation of resonance energies suggests that a more accurate description of the ground state, including correlations beyond the mean-field level, could be essential for reproducing fine details of the experimental strength distribution. Nonetheless, the absence of additional phenomenological parameters enhances the predictive power of the model, establishing it as a valuable framework for future studies of nuclear excitations. In this context, the present study lays the groundwork for the development of fully self-consistent relativistic SRPA formulations. Such extensions could incorporate complex ground-state correlations and phonon couplings, enabling even more accurate modeling of nuclear excitations. 

\begin{acknowledgments}

This research was supported by the Croatian Science Foundation under the project Relativistic Nuclear Many-Body Theory in the Multimessenger Observation Era (HRZZ-IP-2022-10-7773). 
We gratefully acknowledge A. Repko, J. Kvasil and P. Ring for valuable discussions concerning the structure of the excitation spectrum, D. Gambacurta for his suggestions during the initial stages of the development of numerical code, and P. Papakonstantinou for insightful discussions regarding the calculation of the rearrangement terms.

\end{acknowledgments}

\numberwithin{equation}{section}

\appendix

\bibliographystyle{apsrev4-2}
\bibliography{STDA8}
\end{document}